\documentclass[11pt,hidelinks]{article}
\pdfoutput=1
\usepackage[body={17.5cm, 22cm},right=2cm]{geometry}

\usepackage{latexsym,epsfig,amssymb, amsmath,nicefrac}

\usepackage[utf8]{inputenc}

\usepackage{cite}
\usepackage{color}

\usepackage[makeroom]{cancel}
\usepackage{latexsym}
\usepackage{epsf}
\usepackage{amssymb}
\usepackage{graphicx}
\usepackage{amsmath}
\usepackage{amsmath,amssymb,amsthm}
\usepackage{verbatim}
\usepackage[hidelinks]{hyperref}

\usepackage{enumitem}

\newcommand{\eV}{{\, {\rm eV}}}
\newcommand{\keV}{{\, {\rm keV}}}

\newcommand{\GeV}{{\, {\rm GeV}}}

\def\beq{\begin{equation}}
\def\eeq{\end{equation}}
\def\bea{\begin{eqnarray}}
\def\eea{\end{eqnarray}}
\def\bitem{\begin{itemize}}
\def\eitem{\end{itemize}}
\newcommand{\bec}{\begin{center}}
\newcommand{\eec}{\end{center}}
\newcommand{\ba}{\begin{array}}
\newcommand{\ea}{\end{array}}

\def\bar#1{\overline{#1}}

\def\inv{^{\raise.15ex\hbox{${\scriptscriptstyle -}$}\kern-.05em 1}}
\def\lbar{{\lower.35ex\hbox{$\mathchar'26$}\mkern-10mu\lambda}} 

\def\to{\rightarrow}

\let\<=\langle
\let\>=\rangle

\let\+=\uparrow

\def\lsim{\mathrel{\mathop
  {\hbox{\lower0.5ex\hbox{$\sim$}\kern-0.8em\lower-0.7ex\hbox{$<$}}}}}
\def\gsim{\mathrel{\mathop
  {\hbox{\lower0.5ex\hbox{$\sim$}\kern-0.8em\lower-0.7ex\hbox{$>$}}}}}

\begin{document}

\begin{titlepage}
	~\vspace{1cm}
	\begin{center}

	{\Large \bf
		Post-inflationary axions: a minimal target for axion haloscopes
		}

		\vspace{0.7cm}

		{\large
			Marco~Gorghetto$^a$ and
			Edward~Hardy$^b$
			}
		
		\vspace{.6cm}
		{\normalsize { \sl $^{a}$ 
				Department of Particle Physics and Astrophysics, Weizmann Institute of Science,\\
				Herzl St 234, Rehovot 761001, Israel }}
		
		\vspace{.2cm}
		{\normalsize { \sl $^{b}$ Department of Mathematical Sciences, University of Liverpool, \\ Liverpool, L69 7ZL, United Kingdom}}

	\end{center}
	\vspace{1cm}
	\begin{abstract}

An axion-like-particle (ALP) in the post-inflationary scenario with domain wall number $N>1$ can be dark matter if the residual $\mathbb{Z}_N$ symmetry has a small explicit breaking. Although we cannot determine the full dynamics of the system reliably, we provide evidence that such an ALP can account for the observed dark matter abundance while having a relatively small decay constant and consequently a possibly large coupling to photons. In particular, we determine the number of domain walls per Hubble patch around the time when they form using numerical simulations  and combine this with analytic expectations about the subsequent dynamics. We show that the strongest constraint on the decay constant is likely to come from the dark matter ALPs being produced with large isocurvature fluctuations at small spatial scales. We also comment on the uncertainties on the dark matter small-scale structure that might form from these overdensities, in particular pointing out the importance of quantum pressure in the $N=1$ case.

	\end{abstract}

\end{titlepage}

	\thispagestyle{empty}

{\fontsize{11.5}{10.5}
\tableofcontents



\section{Introduction}

By virtue of being the most robust known solution to the strong CP problem, the QCD axion is one of the best-motivated extensions to the Standard Model (SM)~\cite{Peccei:1977hh,Weinberg:1977ma,Wilczek:1977pj,Kim:1979if,Shifman:1979if,Zhitnitsky:1980tq,Dine:1981rt}. It is also plausible that one or more axion-like-particles (ALPs), i.e. pseudo-Nambu-Goldstone bosons of spontaneously broken approximate global U(1) symmetries, exist (we will use the name `axion' to refer to both ALPs and QCD axions). For instance, current understanding suggests that axions might be common in many string theory constructions \cite{Svrcek:2006yi,Arvanitaki:2009fg,Cicoli:2012sz,Acharya:2010zx,Demirtas:2021gsq}.  Additionally, axions are  good candidates to be dark matter (DM) \cite{Preskill:1982cy,Abbott:1982af,Dine:1982ah}: for masses $\lesssim \eV$ and typical interaction strengths, they are stable on cosmological timescales \cite{Arias:2012az,Alonso-Alvarez:2019ssa}, and a nonrelativistic relic population is inevitably produced in the early Universe.

Motivated by these appealing features, numerous axion dark matter detection experiments have been proposed  
 targetting different ranges of axion masses \cite{Sikivie:1983ip,Graham:2015ouw,Irastorza:2018dyq}. Many of these rely on the axion having a coupling to photons
\beq
\mathcal{L} \supset \, \frac14g_{a\gamma\gamma}  a F_{\mu\nu} \tilde{F}^{\mu\nu} ~,
\eeq
where $a$ is the axion, $F_{\mu\nu}$ is the electromagnetic field strength, $\tilde{F}_{\mu\nu}=\frac{1}{2}\epsilon_{\mu\nu\rho\sigma} F^{\rho\sigma}$ is its dual, and  $g_{a\gamma\gamma}$ is the axion-to-photon coupling constant. For the QCD axion this coupling is
\beq \label{eq:ggamma_qcd}
\begin{aligned}
g
_{a\gamma\gamma} &= \frac{\alpha_{em}}{2\pi f_a} \left( \frac{E}{N}-1.92 \right) \\
&= 2.0 \cdot 10^{-16} \GeV^{-1} \left( \frac{E}{N} - 1.92\right) \left(\frac{m_a}{10^{-6}\eV}\right) ~,
\end{aligned}
\eeq
where $f_a$ is the axion decay constant and $\alpha_{em}$ is the fine-structure constant, and the second line follows from the fixed relation between the QCD axion mass and decay constant. $E$ and $N$ are the anomaly coefficients of the Peccei-Quinn (PQ) symmetry with respect to electromagnetism and QCD respectively \cite{Georgi:1986df,GrillidiCortona:2015jxo,Gorghetto:2018ocs}. Meanwhile, the term in eq.~\eqref{eq:ggamma_qcd} that is independent of $E$ comes from the mixing of the axion with the neutral pion, ultimately due to the PQ anomaly with respect to QCD and the fact that quarks are charged. Unless $E$ is very large, the couplings in eq.~\eqref{eq:ggamma_qcd} are challenging to reach experimentally.\footnote{See however e.g. \cite{Farina:2016tgd,Darme:2020gyx,Sokolov:2021ydn} for ways the QCD axion coupling to photons can be enhanced.} 
 This difficulty motivates understanding the plausible values of the axion-to-photon coupling in other theories that are more general than the QCD axion but which might not solve the strong-CP problem.

An ALP's coupling to photons $g_{a\gamma\gamma}$ does not get a contribution analogous to the constant part in eq.~\eqref{eq:ggamma_qcd} (because the U(1) symmetry that gives rise to it is not anomalous under QCD) so might simply vanish. Nevertheless, it is possible that the new physics associated to the ALP carries an anomaly with respect to electromagnetism, so $g_{a\gamma\gamma}$ could plausibly be non-zero. Similarly to eq.~\eqref{eq:ggamma_qcd}, we parametrise such a coupling by
\beq\label{eq:gagg}
g_{a\gamma\gamma} =  \frac{\alpha_{em}}{2\pi f_a} \frac{E}{N} ~. \ 
\eeq
In minimal theories $E$ and $N$ are determined by the anomaly coefficients of the U(1) symmetry with respect to QED and a new confining gauge group, and in many concrete models  $E/N\simeq\mathcal{O}(1)$. 
For convenience we further define the UV model-independent `coupling' $\tilde{g}_{a\gamma\gamma}\equiv g_{a\gamma\gamma}/(E/N)= \alpha_{em}/(2\pi f_a)$; the relation between this and the true coupling ${g}_{a\gamma\gamma}$ depends on the underlying theory. 
Unlike the QCD axion, an ALP's decay constant is independent of its mass. As a result, for fixed values of the mass and $E/N\neq 0$, the axion-to-photon coupling can be large if $f_a$ is small enough. However, if the axion is to comprise the full dark matter abundance in a standard cosmological history, $f_a$ has a lower bound, ultimately because the energy density in the axion field available to produce dark matter is proportional to $f_a^2$.

There are two broad axion cosmological scenarios (see \cite{Redi:2022llj,Harigaya:2022pjd} for an intermediate regime). In the pre-inflationary scenario the global $U(1)$ symmetry that gives rise to the axion is broken during inflation and is never subsequently restored. In this case, the axion field initially takes a constant value over the observable Universe, and the axion relic abundance $\Omega_a$ is produced by the misalignment mechanism. If the ALP's mass is temperature-independent
\beq \label{eq:mis_relic}
\frac{\Omega_a^{\rm mis}}{\Omega_{\rm DM} } \simeq 2.2\cdot10^{-3}   \left(\frac{f_a}{10^{12} ~\GeV} \right)^2  \left(\frac{m_a}{10^{-6} ~\eV} \right)^{1/2} h(\theta_0)\,\theta_0^{2}  ~,
\eeq
where $\Omega_{\rm DM}$ is the observed DM abundance, $\theta_0 \in (-\pi,\pi]$ is the initial misalignment angle,  and $h$ is a function that accounts for anharmonicities in the axion potential. For typical axion potentials $h\left(\theta_0\right)\simeq 1$ for $\theta_0\lesssim 1$ and 
 increases slowly as $\theta_0$ approaches $\pi$.  Constraints on the size of primordial isocurvature perturbations give a lower bound on $\pi-|\theta_0|$ that depends on the Hubble scale during inflation, see Appendix~\ref{aa:isopre}. 
 For instance, for  a cosine potential $h(\theta_0)$ increases only logarithmically with $1/(\pi-|\theta_0|)$, and $H_I\simeq 1\eV$ requires $\pi-|\theta_0| \gtrsim 10^{-18}$ which gives $h(\pi- 10^{-18}) \lesssim 200$. 
 From eq.~\eqref{eq:mis_relic}, if $\Omega^{\rm mis}_a=\Omega_{\rm DM}$ the axion-to-photon coupling is constrained to be~\footnote{An ALP's misalignment relic abundance, and consequently the maximum  $\tilde{g}_{a\gamma\gamma}$, can be larger if the axion mass is temperature dependent. This is also possible in theories with multiple axions \cite{Cyncynates:2021xzw}, different cosmological histories, and different axion potentials  \cite{Alonso-Alvarez:2017hsz}.}
\beq \label{eq:mis}
\tilde{g}_{a\gamma\gamma} \lesssim  5\cdot 10^{-17}\GeV^{-1}\left(\frac{m_a}{10^{-6}\eV} \right)^{1/4} h(\theta_0)^{1/2}\,\theta_0 ~.
\eeq

In the post-inflationary scenario the $U(1)$ symmetry is instead unbroken at the end of inflation or is subsequently restored e.g. by finite temperature effects. As a result, the field is initially inhomogeneous over our observable Hubble patch and the axion relic abundance is determined by the decay of a network of topological defects: global cosmic strings and domain walls. 
In this paper, we argue that an ALP in the post-inflationary scenario can be all the DM with a much smaller $f_a$ (i.e., for  fixed $E/N$, a larger coupling to photons) than in the pre-inflationary scenario while being compatible with observational constraints, the most important of which is expected to come from bounds on primordial isocurvature perturbations.\footnote{A related analysis of the same theory can be found in \cite{Gelmini:2021yzu,Gelmini:2022nim}  focusing on the gravitational waves from domain walls, see also \cite{ZambujalFerreira:2021cte} for a similar study of a different theory that has an axion.}  This is possible because, if $N>1$, the string-domain wall network might be long-lived, which results in the dark matter produced by its decay being redshifted less.  We also analyse the impact of the isocurvature constraints on possible gravitational wave signals from topological defects and discuss the properties of, and the uncertainties in, the dark matter substructure that might form in the early Universe.

We stress that it is difficult to analyse the system's dynamics from first principles because they are both non-linear and involve large scale separations, and as a result the field evolution is presently unknown. 
 Nevertheless, we make progress by combing results from numerical simulations with reasonable standard assumptions about the dynamics of the domain wall network (which we can only very partially check). 
In particular, {in Section~\ref{sec:sims}} we determine number of domain walls per Hubble patch $\mathcal{A}_\star$ at the time when the axion mass becomes cosmologically relevant (when the Hubble parameter is $H_\star$), which is important for the resulting dark matter abundance{, and provide evidence that the axions emitted from the string-domain wall system are, at most, mildly relativistic}. Our results for $\mathcal{A}_\star$, applicable for both ALPs and the QCD axion, suggest that this quantity increases logarithmically with $f_a/H_\star$, and mean that an extrapolation of simulation results is needed. We also discuss the numerous remaining uncertainties on the evolution of $\mathcal{A}$ after this time, and on the spatial distribution of the axions produced from the string-domain wall system.

The structure of the paper  is as follows. In Section~\ref{sec:theory} we introduce the post-inflationary scenario for ALPs and discuss analytic expectations about the dynamics and the dark matter relic abundance, and 
in Section~\ref{sec:sims} we present results from numerical simulations of the string-wall system. In Section~\ref{sec:signals} we analyse the observational constraints on the theory and discuss its relevance for upcoming axion dark matter searches, and we also comment on the possible gravitational waves signals and dark matter substructure. In Section~\ref{sec:conclusions} we describe future directions and extensions.

\section{Early universe evolution}\label{sec:theory}

An axion-like-particle (ALP) is the pseudo-Nambu Goldstone boson, $a$, of an approximate global U(1) symmetry that is spontaneously broken at a high scale $v$.  Because $a$ is an angular variable, the theory preserves an exact discrete shift symmetry $a\to a+2\pi v$. The simplest model in which this happens consists of a complex scalar $\phi$ with an approximate global U(1) symmetry
and potential 
\begin{equation}\label{eq:Vphi}
V_\phi(\phi)=\frac{m_r^2}{2v^2}\left(|\phi|^2-v^2\right)^2+\dots ~,
\end{equation}
that induces a vacuum expectation value $|\langle\phi\rangle|=v$.  Expanding $\phi=\frac{1}{\sqrt{2}}(r+v)e^{ia/v}$, the axion $a$ is a Goldstone mode, while the radial mode $r$ has (large) mass $m_r$, which is often of order $v$.

We assume that the U(1) symmetry is broken by non-perturbative effects, e.g. because it is anomalous with respect to a new confining gauge group. Such a breaking leads to a potential for $a$ that is invariant under $a\to a+2\pi f_a$. In general $f_a$ and $v$ are related by $v=N f_a$, where $N$ has to be an integer otherwise the symmetry $a\to a+2\pi v$ would be violated (in KSVZ-like models, $N$ is the number of fermions charged under U(1)). As a result, the potential induced by a particular non-perturbative contribution does not necessarily break the U(1) symmetry completely, and can leave a subgroup $\mathbb{Z}_N$ unbroken. 
For instance, the total axion potential might be
\begin{equation} \label{eq:ax_pot}
	V(a)= m_a^2 f_a^2 \left[1- \cos (a/f_a)\right] + \delta V_{PQ}(a)  ~,
\end{equation}
where the first term comes from the non-perturbative contribution. 

In eq.~\eqref{eq:ax_pot}, $\delta V_{PQ}(a)$ represents additional sources of U(1) breaking that violate the discrete symmetry $a\to a+2\pi f_a$ while still preserving $a\to a+2\pi v$. Such breaking is expected because it is believed that there are no exact global symmetries in quantum gravity  \cite{Abbott:1989jw,Banks:2010zn,Harlow:2018tng} (implications for the QCD axion have been studied in e.g.  \cite{Kamionkowski:1992mf,Barr:1992qq,Ghigna:1992iv,Rai:1992xw,Holman:1992us,Dine:1992vx,Dobrescu:1996jp,Cox:2019rro,Fichet:2019ugl,Yin:2020dfn}).\footnote{This happens e.g. via higher dimensional operators of dimension $n+m$ of the form $\alpha\phi^n (\phi^{\dagger})^m/M_{p}^{n+m-4}$, where $M_{p}$ is the Planck Mass and $n\neq m$. If $\alpha=\mathcal{O}(1)$, $n+m$ needs to be large in order for the U(1) to be an approximate symmetry. On the other hand, if $\alpha$ is exponentially suppressed (as expected from some non-perturbative gravitational processes such as wormholes~\cite{Abbott:1989jw,Banks:2010zn}) $n+m$ can be small.} As an example, although the particular form does not matter, {one can consider}
\begin{equation} \label{eq:ax_pot2}
\delta V_{PQ}(a)=\epsilon f_a^4 \cos(a/(N f_a)-\theta_1) ~,
\end{equation}
with  $\theta_1=\mathcal{O}(1)$. In the absence of $\delta V_{PQ}$ there would be $N$ inequivalent degenerate vacua, but the energy densities of these vacua are shifted by $\simeq \epsilon f_a^4$ due to by $\delta V_{PQ}$. 
We demand $\epsilon<m_a^2/f_a^2$ so that the global minimum of the full potential is a perturbation of one of the $\mathbb{Z}_N$ preserving minima and the axion mass is dominated by the first term in eq.~\eqref{eq:ax_pot}~\cite{Banerjee:2022wzk}. In the following we will take $\theta_1=0$, but this choice will not have any effect.

The equations of motion (EoM) from the potentials in eqs.~\eqref{eq:Vphi} and~\eqref{eq:ax_pot} admit topologically non-trivial field configurations called strings and domain walls respectively \cite{Kibble:1976sj,Vilenkin:1981kz,Vilenkin:1982ks,Sikivie:1982qv} (see \cite{Sikivie:2006ni} for a review). These form if the field is inhomogeneous, as is the case in the post-inflationary scenario.   
We assume a standard cosmological history, and, as we discuss subsequently, in all theories of interest the strings and domain walls are destroyed before matter-radiation equality (MRE). 
During the preceding era of radiation domination the Hubble parameter is  $H \equiv \dot{R}/R=1/(2t)$, where $R$ is the scale factor of the Universe and $t$ is the cosmic time.  Unless otherwise stated we assume that the axion potential is independent of the temperature of the Universe.

\subsection{The scaling regime} \label{ss:post_dynamics}

After the PQ symmetry spontaneously breaks, $\phi$ has initially random spatial fluctuations. The axion potential in eq.~\eqref{eq:ax_pot} is cosmologically irrelevant while  $H \gg m_a$ so a network of cosmic strings forms.  
The strings occur in regions of space in which the axion U(1) is wrapped non-trivially, and they have (evolving) energy per unit length, i.e. tension, 
\beq \label{eq:mu}
\mu= \pi v^2 \log(m_r/H)~.
\eeq
The string network is driven to an attractor, scaling, solution on which there is approximately one Hubble length of string per Hubble volume.  
To study this regime, numerical simulations of the nonlinear EoM from the potential in eq.~\eqref{eq:Vphi} are required and an extrapolation in $\log(m_r/H)$ is needed. In particular, simulations can only be carried out with a relatively small value of $\log(m_r/H) \lesssim 8\div9$ so the scale separation between the string thickness (of order $m_r^{-1}$) and the Hubble distance is far from the values relevant shortly before $H=m_a$~\cite{Gorghetto:2018myk}.  

We define the number of strings per Hubble volume $\xi= \lim_{\mathcal{V}\to \infty} \ell_{\rm st} t^2/\mathcal{V}$ where $\ell_{\rm st}$ is the length of string in a volume $\mathcal{V}$.  
Simulations show that during the scaling regime $\xi$ has a logarithmic increase, $\xi \simeq 0.24 \log(m_r/H)$. 
The energy density in (long) strings is, from eq.~\eqref{eq:mu}, $\rho_{\rm s}  = 4\xi \mu H^2$ 
and the network releases energy density into (relativistic) axion waves at a rate $\Gamma_a^{\rm st}\simeq \xi\mu/t^3$. Additionally, simulations suggest that at the relevant large  $\log(m_r/H)$ most of the axions are emitted with momentum of order Hubble. The resulting axion number density released until time $t$ is given by $n_a^{\rm st}(t)\simeq\xi \mu H$, which is dominantly produced in the final few Hubble times prior to $t$, see Appendix~\ref{app:st}.\footnote{See however  \cite{Buschmann:2021sdq} which suggests a scale invariant instantaneous emission spectrum, which would reduce the relic abundance by a factor of $\log_\star$ relative to eq.~\eqref{eq:string_relic}.}
  
 The scaling regime continues until the time when the Hubble parameter $H$ equals the axion mass, $H=m_a$, when the first ($\mathbb{Z}_N$-preserving) contribution to the axion potential in eq.~\eqref{eq:ax_pot} starts to be cosmologically relevant. We indicate quantities at that moment by $\star$, e.g. $H_\star$ is the Hubble parameter at this time, and we define  $\log_\star \equiv \log(m_r/H_\star)\simeq\log(f_a/m_a)$, which is of $\mathcal{O}(100)$ for the $(m_a,f_a)$ that we will consider.

 The axion waves emitted  by strings prior to $H_\star$ become nonrelativistic after $H=H_\star$ and  contribute to the dark matter relic abundance,  see Section~4.1 of~\cite{Gorghetto:2021fsn} and \cite{Gorghetto:2020qws} for more details.   Evaluating $n_a^{\rm st}$ at $H=H_\star$ and redshifting to today, the resulting dark matter from strings during the scaling regime, $\Omega_a^{\rm st}$, is, for a temperature-independent $m_a$,
 \beq \label{eq:string_relic}
 \frac{\Omega_a^{\rm st}}{\Omega_{\rm DM}} \simeq 3  \left( \frac{\xi_\star \log_\star}{10^3} \right) \left(\frac{N f_a}{ 10^{12} ~\GeV} \right)^2 \left(\frac{m_a}{10^{-6} ~\eV} \right)^{1/2}  ~.
 \eeq
For the axion masses and decay constants of interest, $10^{-15} ~\eV \lesssim m_a \lesssim \eV$ and $10^{8}~\GeV \lesssim f_a \lesssim 10^{12}~\GeV$, we have $\log_\star\simeq 40 \div 80$. Therefore, assuming that the logarithmic growth in $\xi$ is valid also for $\log(m_r/H)\gg1$,  $\xi_\star \log_\star \simeq 10^3$ (if $\xi_\star$ take a different value the abundance can be calculated accordingly from eq.~\eqref{eq:string_relic}). Note that for larger values of $\xi_\star \log_\star \gtrsim 10^3$ or $N\gg1$, the axion number density is substantially reduced  between the time when the axions are relativistic and when they become nonrelativistic, because the amplitude of the axion field is much larger than $f_a$ and nonlinear effects from the axion potential become relevant. As a result, the abundance $\Omega_a^{\rm st}$ decreases with respect to  eq.~\eqref{eq:string_relic}, see~\cite{Gorghetto:2021fsn}. For $\xi_\star \log_\star \lesssim 10^3$ the effect is of order 1, comparable to the other uncertainties in eq.~\eqref{eq:string_relic}.\footnote{The non-linear evolution is more important  for the QCD axion because $m_a$ has a strong temperature dependence \cite{Gorghetto:2020qws}.}

\subsection{The expected post-inflationary relic abundance} \label{ss:relic}

Once the axion potential is relevant at $H=m_a= H_\star$, domain walls form, across which the axion field interpolates between vacua. Each string is bounded by $N$ domain walls. The domain walls have a tension, i.e. energy per unit area, given by $\sigma = \beta m_a f_a^2$, where $\beta$ is a numerical factor ($\beta=8$ for a cosine and is similar for other typical potentials). 

\subsubsection*{\pmb{$N=1$}}

If $N=1$ the domain walls connect the same vacuum of the potential in eq.~\eqref{eq:ax_pot}, $a/f_a=0$, and the string and domain wall network is unstable and expected to decay. In the process axion waves are emitted, and subsequently the axion field oscillates around the global minimum of its potential and behaves as dark matter. The contribution to the axion abundance produced by the domain walls and strings at this time is hard to determine reliably. If the network is destroyed soon after $H=H_\star$, it might be relatively unimportant compared to that from the scaling regime, $\Omega_a^{\rm st}$ in eq.~\eqref{eq:string_relic},  because the energy in strings is enhanced relative to that in domain walls by the logarithmically boosted string tension (see eq.~\eqref{eq:naw2} below).  Including only the contribution from the scaling regime, the value of $f_a$ for which an $N=1$ post-inflationary ALP makes up all the DM is of same order as that in the pre-inflationary scenario with an initial misalignment angle $\theta_0\sim \pi-10^{-18}$ close to the top of the potential, see  eq.~\eqref{eq:mis} and Figure~\ref{fig:signal}. We do however stress that eq.~\eqref{eq:string_relic} is a lower bound on the $N=1$ axion relic abundance; the domain wall network might take longer to decay than expected, in which case it could produce a sizable additional contribution resulting in smaller required $f_a$.

\subsubsection*{\pmb{$N>1$}}

In contrast, if $N>1$ the domain walls interpolate between the inequivalent vacua of the first term in eq.~\eqref{eq:ax_pot}. In the absence of $\delta V_{PQ}$ they would form a topologically stable network after $H=H_\star$ \cite{Sikivie:1982qv} which would come to dominate the energy density in the Universe~\cite{Zeldovich:1974uw}.  
However, the  energy bias between the vacua induced by $\delta V_{PQ}$ is believed to lead to the destruction of the network when $\delta V_{PQ}$ becomes cosmologically relevant \cite{Sikivie:1982qv,Gelmini:1988sf,Larsson:1996sp,Ringwald:2015dsf}. This is thought to happen when the energy density in domain walls, $\rho_{\rm w}\simeq\sigma H$, is approximately equal to the energy density provided by the bias term, $\delta V_{PQ}\simeq \epsilon f_a^4$, i.e. the Hubble parameter when the network is destroyed is $H_d \simeq \epsilon f_a^2/m_a$. The energy released by the domain walls is expected (but not proven) to dominantly go into dark matter axions, with a small fraction emitted as gravitational waves \cite{Gelmini:2021yzu}.

If $\epsilon\simeq m_a^2/f_a^2$, the energy density difference between the local minima is of $\mathcal{O}(m_a^2f_a^2)$ and the $\mathbb{Z}_N$ symmetry is badly broken. As a result, the evolution of the string-domain wall network, and the axion relic abundance, is likely to be similar to that in the $N=1$ scenario.  
However, if $\epsilon \ll m_a^2/f_a^2$ the domain wall network is long-lived and is destroyed when the Hubble parameter is $H_d \ll m_a$. Unfortunately, the dynamics in this case are even harder to study with numerical simulations than the $N=1$ scenario. This is because there is another relevant scale, $m_a$, so a second extrapolation, in $\log(m_a/H)$, is needed. Simulations can only reach values $\log(m_a/H) \lesssim 5$, whereas we will see in Section~\ref{ss:constraints} that values up to $\log(m_a/H_d)\simeq 60$ (determined by $\delta V_{PQ}$) are possible. The too-small $\log(m_r/H)$ are an issue not only because the string tension depends on this factor, but also because the string dynamics are dominated by emission of high energy modes with energy of order $m_r$ for $\log(m_r/H)\lesssim 9$~\cite{Gorghetto:2020qws}. The second extrapolation is perhaps even more problematic because, as $m_a/H$ increases, there is a growing mismatch between the typical spatial size of the domain walls, which extend over an Hubble length $\sim H^{-1}$, and the frequency of the axions they are expected to emit {(of order $m_a$)}. It is plausible that this could e.g. decrease the ability of the domain walls to radiate axions as $m_a/H$ increases, which might even qualitatively change the properties of the network when $\log(m_a/H)\gg1$. 

Nevertheless, we proceed with some standard analytic expectations, which we will subsequently partially check. It is useful to define the domain-wall area per Hubble patch $\mathcal{A} = \lim_{\mathcal{V}\to\infty} A t/\mathcal{V}$, where $A$ is the total domain wall area in a volume $\mathcal{V}$. $\mathcal{A}$  measures parametrically the total area of domain walls in one Hubble patch in units of the Hubble area $H^{-2}$, so the energy density in domain walls is 
\beq
\begin{aligned}
	\rho_{\rm w}  = 2 \mathcal{A}\sigma H ~. 
\end{aligned}
\eeq
While $\delta V_{PQ}$ is cosmologically irrelevant, at $H<H_\star$ the domain walls are thought to survive in a domain wall `scaling regime': $\mathcal{A}$ is expected to be approximately constant in time, and likely not far from $\mathcal{A}\simeq \mathcal{O}(1)$ \cite{Hiramatsu:2012sc} (or slightly larger, because of the logarithmic violation discussed in Section~\ref{sec:sims}). To maintain this, the network must emit energy density at a rate 
\beq \label{eq:GammaW}
\Gamma_a^{\rm w} \simeq  \rho_{\rm w}/(2t)~,
\eeq
i.e. an energy density of order $\rho_{\rm w}(t)$ is released every Hubble time.\footnote{This is because $\Gamma_a^{\rm w}(t)=[\dot{\rho}^{\rm free}_{\rm w}(t_0)-\dot{\rho}_{\rm w}(t_0)]_{t_0=t}$, where $\rho^{\rm free}_{\rm w}(t)=\mathcal{A}\sigma/\sqrt{t_0t}\propto R(t)^{-1}$ is the energy density of a system of domain walls that coincides with that in the scaling regime at $t=t_0$, but evolves freely,  i.e. the domain walls are fixed in comoving coordinates. } 
The momentum of the axions produced is unknown and could be set by either of the two scales relevant to the domain walls: their size $\simeq H^{-1}$ or their thickness $\simeq m_a^{-1}$. Although the latter is suggested by results of simulations at small $\log(m_a/H_d)$ and $\log(m_r/H_d)$, see Section~\ref{sec:sims}, it is not clear whether this remains the case for physical parameters. Regardless of this uncertainty, because the emitted energy dominantly goes into at most semi-relativistic modes, the axion number density produced until time $t$ is approximately $m_a^{-1}\int^t_{t_\star}dt'\Gamma(t')(R(t')/R(t))^3$, i.e. 
\begin{align}\label{eq:naw1}
	n_a^{\rm w}(t) & \simeq \int_{\log t_\star}^{\log t} \mathcal{A}  \frac{\beta f_a^2}{2t'}  \left(\frac{R(t')}{R(t)} \right)^3 ~d\log t'  \\
	& 	=	2\mathcal{A}\beta  m_af_a^2\left[ \left(\frac{m_a}{H}\right)^\frac12 -1 \right]\left(\frac{R_\star}{R(t)} \right)^3\label{eq:naw2}
\end{align}
where $R(t)\propto t^{1/2}$ is the scale factor. Because $R(t')^3/t' \propto t'^{1/2}$, it is clear from eq.~\eqref{eq:naw1} that the number density is mostly produced at the latest times that the network survives to and is dominated by the axions emitted during the Hubble time immediately preceding the domain wall network's destruction. If the network survives up to $H\ll m_a$, the axion number density is parametrically enhanced, by a factor $(m_a/H)^{1/2}$, with respect to that from misalignment, which is $\mathcal{O}(m_af_a^2)$ at $t=t_\star$, see eq.~\eqref{eq:naw2}.

At $H=H_d$, all the domain walls are destroyed and their energy density $\rho_{\rm w}= 2 \mathcal{A}_d\sigma H_d$ is released, where $\mathcal{A}_d$ denotes the value of $\mathcal{A}$ at this time. The resulting axion number density is similar that left from the evolution of domain walls up to $H=H_d$, calculated in eq.~\eqref{eq:naw2}.  
Therefore, the axion relic abundance from energy stored in domain walls is approximately
\beq
\begin{aligned} \label{eq:relic}
	\frac{\Omega_a^{\rm w}}{\Omega_{\rm DM}} & \simeq \frac{ \beta \mathcal{A}_d m_a f_a^2}{T_{\rm eq} H_d^{1/2} M_{\rm Pl}^{3/2}} \\ 
	&\simeq 0.1 \mathcal{A}_d  \left(\frac{m_a }{H_d} \right)^{1/2}  \left(\frac{f_a}{ 10^{12} ~\GeV} \right)^2  \left(\frac{m_a}{10^{-6} ~\eV} \right)^{1/2} ~,
\end{aligned}
\eeq
where we set $\beta= 10$ in the second line. For $H_d \ll m_a$ this contribution far exceeds that produced by strings before $H=H_\star$ given in eq.~\eqref{eq:string_relic}. Moreover, in Appendix~\ref{app:sim} we show that the abundance produced from energy in strings after $H=H_\star$ is in turn smaller than that released by strings before $H=H_\star$ (fundamentally because $\rho_s$ and therefore $\Gamma_s$ decrease faster than $\rho_{\rm w}$ and $\Gamma_{\rm w}$ as $t$ increases). Therefore we expect that eq.~\eqref{eq:relic} approximates the total relic abundance in this scenario.

If $H_d \ll m_a$ (i.e. $\epsilon$ is small enough), an axion in the post-inflationary scenario with $N>1$ can account for the full dark matter abundance with smaller $f_a$ than in the $N=1$ post-inflationary and the pre-inflationary scenario for any value of $\theta_0$ compatible with isocurvature constraints. The possible coupling of such an axion with $E/N\simeq 1$ to photons is correspondingly enhanced, {in particular} 
\beq \label{eq:gaggD}
\tilde{g}_{a\gamma\gamma} \simeq 5\cdot 10^{-16}\GeV^{-1}  \mathcal{A}_d^{1/2} \left(\frac{m_a}{H_d} \right)^{1/4}  \left(\frac{m_a}{10^{-6} ~\eV} \right)^{1/4}~.
\eeq
 As mentioned, we expect $H_d\sim \epsilon f_a^2/m_a$, but we keep eq.~\eqref{eq:gaggD}  in terms of $H_d$ because we do not have a reliable way of predicting the relation between $H_d$ and $\epsilon$.

\subsection{The density power spectrum}\label{ss:power}

The number of strings and domain walls per Hubble patch varies by order one factors  between different Hubble patches. Therefore, immediately after the decay of the string-wall network (when most of the axions are expected to be produced), the energy density of the axion field is expected to be left with substantial inhomogeneities over Hubble distances.  Being only in the axion energy density, these are isocurvature fluctuations. Their properties are conveniently described by the power spectrum $\mathcal{P}_\delta(k)$ of the overdensity field $\delta(\vec{x})\equiv (\rho_a(\vec{x})-\bar{\rho}_a)/\bar{\rho}_a$, where $\rho_a=\frac12\dot{a}^2+\frac12m_a^2a^2$ is the non-relativistic energy density of the axion when the field is close to its minimum, $\bar{\rho}_a$ is its spatial average, and we defined
\beq \label{eq:pow}
\langle \tilde{X}^*(t,\vec{k})\tilde{X}(t,\vec{k}')\rangle\equiv (2\pi)^3\frac{2\pi^2}{k^3}\delta^3(\vec{k}-\vec{k}') \mathcal{P}_X(t,k) ~,
\eeq
for a generic field $X$ with Fourier transform $\tilde{X}=\int d^3x\exp(-i\vec{k}\cdot \vec{x})X(\vec{x})$. Unfortunately the power spectrum of a post-inflationary ALP is not known; below we summarise its possible form based on analytical arguments and hints from numerical simulations.

\begin{figure*}[t]
	\begin{center}
		\includegraphics[width=0.5\textwidth]{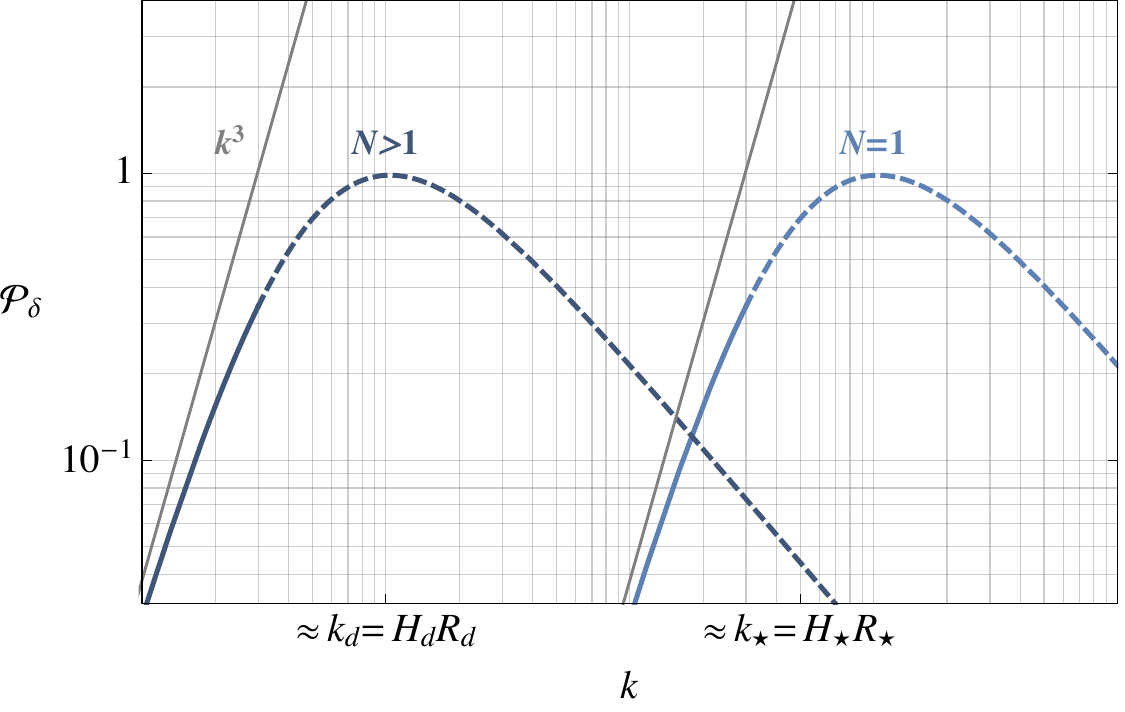}
	\end{center}
	\caption{
The schematic expected form of the primordial power spectrum $\mathcal{P}_\delta(k)$ of the axion over-density field $\delta=(\rho_a-\bar{\rho}_a)/\bar{\rho}_a$ after the string-domain wall network decays for $N=1$ and $N>1$
, where $k$ is the comoving momentum. 
$\mathcal{P}_\delta(k)$ is thought to be peaked at a scale set by the Hubble parameter when the network decays and to have a `white noise' $k^3$ tail at smaller momenta. $H_d$ and $R_d$ are the Hubble parameter and scale factor when the network is destroyed and $H_\star$ and $R_\star$ are those when $H=m_a$. 
Free-streaming could suppress the peak of $\mathcal{P}_\delta$ in both scenarios if the dark matter axions are produced mildly relativistic, and we plot $P_\delta$ dashed in this region to emphasize the especially large uncertainty.
\label{fig:cartoon}}
\end{figure*}

If $N=1$, the network decays at $H\simeq H_\star$ and therefore the resulting fluctuations are likely to be on comoving scales approximately set by $k_\star \equiv H_\star R_\star$. Numerical simulations at small scale separation, $\log(m_r/H_\star)\simeq5\div6$, support this picture, with $\mathcal{P}_\delta(k)$ peaked (and obtaining an $\mathcal{O}(1)$ value) at comoving momenta $k\simeq \mathcal{O}(10) k_\star$ \cite{Vaquero:2018tib,OHare:2021zrq,Gorghetto:2021fsn}. (This corresponds to $\delta$ of order one at spatial scales approximately set by $\lambda_\star/\mathcal{O}(10)$, with $\lambda_\star=2\pi/k_\star$.) At larger comoving scales, $k\ll k_\star$, the perturbations are uncorrelated, by causality. From eq.~\eqref{eq:pow}, this fixes the spectrum to have the `white-noise' form $\mathcal{P}_\delta(k)=  \left(C k/k_\star\right)^3$ for $k\lesssim k_\star/C$, where $C$ is a numerical constant.  
The axion waves are expected to be produced marginally relativistic (with momentum $\mathcal{O}(10) k_\star$, i.e. wavelength coinciding with the typical size of the strings and domain walls -- which might also be connected to the density spectrum being peaked at $\simeq 10k_\star$). Therefore they initially free stream over comoving distances $\simeq (H_\star R_\star)^{-1}$. This might partly wash out the order-one fluctuations, but should leave the IR ($k^3$) part of the spectrum largely unchanged because the axions will not free-stream over the corresponding distances~\cite{OHare:2021zrq}. In Figure~\ref{fig:cartoon} we show a sketch of the anticipated power spectrum.

If $N>1$ the spatial distribution of domain walls is expected to lead to order one fluctuations in the axion energy density immediately after the network's destruction. Although not yet established with simulations, it is plausible that these will be on scales set parametrically by $k_d \equiv H_d R_d$, where $\mathcal{P}_\delta$ will be correspondingly peaked, i.e. on comoving spatial scales that are larger than $k_\star^{-1}$. Similarly to the $N=1$ case, at $k\lesssim k_d$ the spectrum acquires a $k^3$ white-noise form
\beq \label{eq:Pdelta}
\mathcal{P}_\delta(k) \simeq 
(C k/k_d)^3  \text{  for $k\lesssim k_d/C$} 
~,
\eeq
where $C$ is an undetermined constant, possibly not far from order one. As discussed in the previous Section, the axion momentum distribution is uncertain in this scenario. If axions are dominantly produced with momentum $H_d$, they are initially highly non-relativistic for $H_d\ll m_a$ and their free-streaming will be  irrelevant. However, if they are produced with momentum $m_a$, they are initially semi-relativistic and will free-stream over comoving distances of order $(H_d R_d)^{-1}$.\footnote{More precisely, if produced with momentum $m_a$ the axions would free-stream a comoving distance $\simeq H_d^{-1} R_d^{-1} (1+ \log(t_{\rm eq}/t_d))$ where $t_{\rm eq}$ is the cosmic time at matter-radiation equality \cite{Kolb:1990vq}.} Similarly to the $N=1$ scenario, fluctuations on comoving scales $k \gtrsim k_d$ will be affected, but the $k^3$ part of $\mathcal{P}_\delta$ (which is what will be relevant to the constraints that we discuss shortly) is expected to be left unchanged. Unfortunately, the reach in $\log(m_a/H_d)$ and $\log(m_r/H_\star)$ in simulations is not sufficient for us to extrapolate to obtain the value of $C$ with the physical parameters and  the details of the peak of  $\mathcal{P}_\delta$ are especially uncertain.

\section{Comparison with numerical results} \label{sec:sims}

\subsection{The area parameter}

The  area parameters $\mathcal{A}_\star$ and $\mathcal{A}_d$ {entering in the abundance of eq.~\eqref{eq:relic}} can be extracted from numerical simulations of eqs.~\eqref{eq:Vphi} and~\eqref{eq:ax_pot}, although  these have to be extrapolated in $\log(m_r/H)$ and, for $\mathcal{A}_d$, also $\log(m_a/H)$. 
More precisely, we define $\mathcal{A}$ to be the area of the surfaces where the axion is at the maxima of the $\mathbb{Z}_N$-preserving potential in eq.~\eqref{eq:ax_pot}, $a/f_a=2\pi n$, $n=0,\dots,N-1$, which correspond to domain walls when $H\lesssim m_a$. Details of the simulations and additional results may be found in Appendix~\ref{app:sim}.

\begin{figure}[t]
	\begin{center}
		\includegraphics[width=0.47\textwidth]{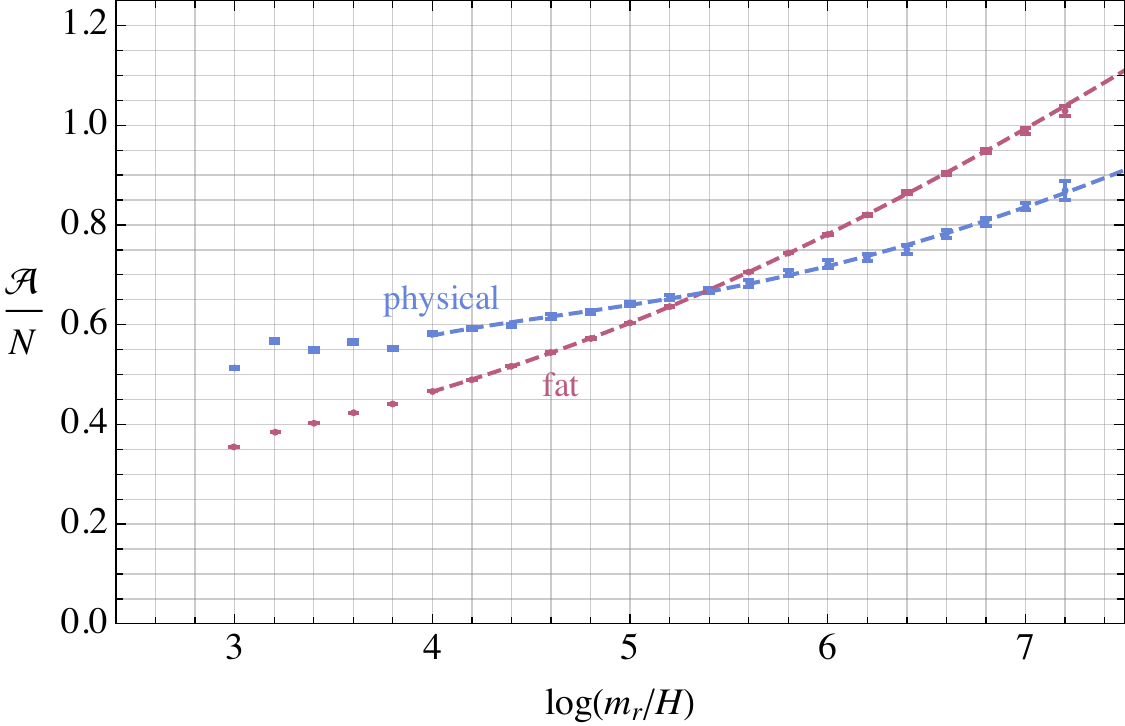} \ \ \ \  
		\includegraphics[width=0.47\textwidth]{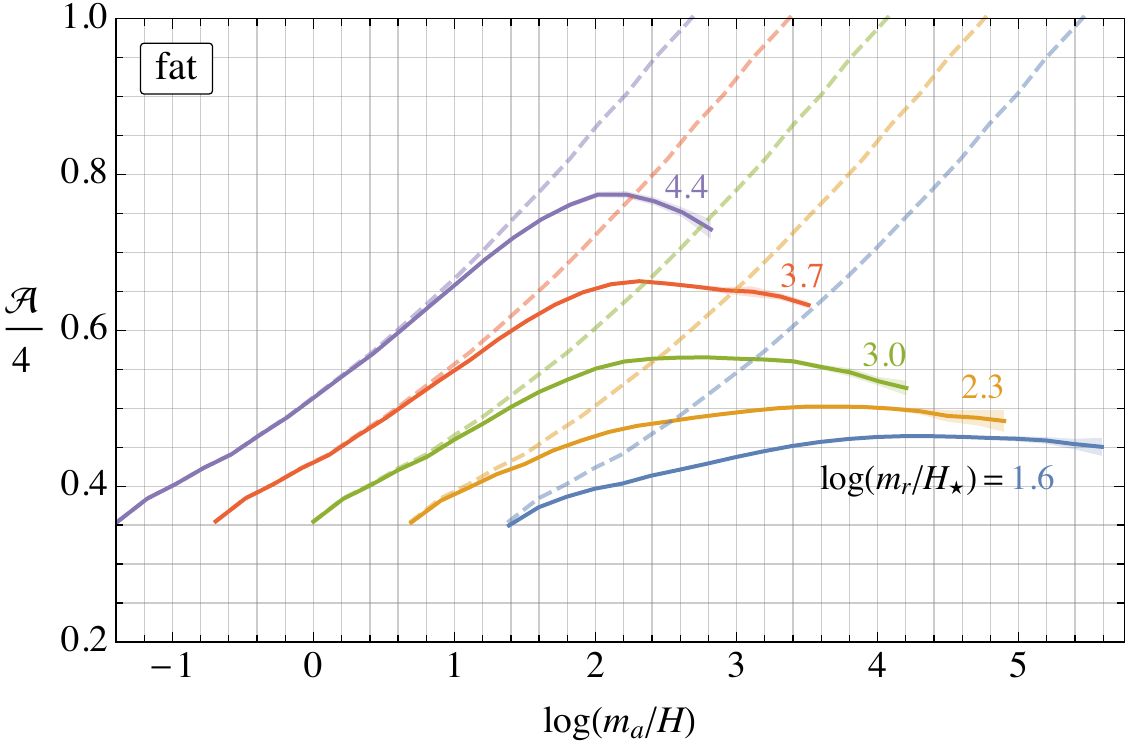} 
	\end{center}
	\caption{{\bf \emph{Left:}}
		The time evolution of $\mathcal{A}$, which measures the area of the surfaces on which $a/f_a=2\pi n$, $n=0,\dots,N-1,$ where $N$ is the domain wall number, during the string scaling regime (i.e. with $m_a=0$) in fat and physical theories. Time is represented by $\log(m_r/H)$, and we factor out $N$ because $\mathcal{A}\propto N$. A clear increase of $\mathcal{A}$ with $\log(m_r/H)$ is evident over a few e-foldings. We also plot the best fits to the asymptotically-linear function in eq.~\eqref{eq:A_linear}.  		
		{\bf \emph{Right:}} The evolution of $\mathcal{A}$ with time, parameterised by $\log(m_a/H)$, for different values of the axion mass (i.e. different $\log(m_r/m_a)\equiv\log(m_r/H_\star)$) and no explicit $\mathbb{Z}_N$ breaking for the fat string system with $N=4$. For comparison we also show results from simulations with $m_a=0$, dashed. As expected, when $m_a\ll H$ the evolution matches that with $m_a=0$. At $H\lesssim m_a$ the surfaces measured by $\mathcal{A}$ are domain walls. The increase in $\mathcal{A}$ while $m_a\ll H$  results in a larger domain wall area at $H\simeq m_a$ for larger $\log(m_r/H_\star)$. At $H\lesssim m_a$ the growth in $\mathcal{A}$ ends, although its asymptotic behaviour is challenging to pin down.
		\label{fig:arasim}}
\end{figure} 

Figure~\ref{fig:arasim} (left) shows the evolution of $\mathcal{A}$  with $m_a=0$, i.e. during the string scaling regime, with time parametrised by $\log(m_r/H)$. In this era $\mathcal{A}$ is proportional to $N$, so we plot  $\mathcal{A}/N$. Results are given for both the so-called `fat' string system (in which $m_r\propto R^{-1}$ in eq.~\eqref{eq:Vphi}) and the physical system (in which $m_r$ is constant), with initial conditions close to the attractor solution.\footnote{The fat string system displays better numerical properties. For instance, because more cosmic time is available between two fixed values of $m_r/H$, this system reaches the attractor solution faster.}  
Not surprisingly given the logarithmic increase of $\xi$ on the attractor, $\mathcal{A}$ grows with  $\log(m_r/H)$.  For the fat string system $\mathcal{A}$ is approximately proportional to $\xi$; this can be seen from the plot of $\mathcal{A}(t)/\xi(t)$ in Figure~\ref{fig:scal} in Appendix~\ref{aa:scal}. The results for $\mathcal{A}(t)/\xi(t)$ for the physical system are less clear but are compatible with this ratio approaching a constant value. This possibility is supported by the fact that $\mathcal{A} \propto \xi$ is automatic if the average string curvature is proportional to the Hubble parameter, as is expected by causality during the scaling regime. On the other hand, even for the fat string system, the uncertainty on $\mathcal{A}(t)/\xi(t)$ is still large enough that we cannot exclude a deviation at much later times, $\log(m_r/H)\gg1$.

Given that  $\xi(t)$ during the scaling regime is well fitted by a linear function of $\log\equiv\log(m_r/H)$ \cite{Gorghetto:2020qws}, a natural ansatz for  $\mathcal{A}$ is 
\begin{equation}\label{eq:A_linear}
\mathcal{A}(t)/N=c_1\log+c_0+c_{-1}\log^{-1} + c_{-2}\log^{-2} \, ,
\end{equation}
where the last two terms encode possible deviations from linearity that are irrelevant at late times, but affect the evolution at the small times (i.e. $\log$s) that are accessible to simulations.  
This form fits the fat string data well with $c_1=0.35(2)$. It also fits the physical data well, however the value of $c_1$ is more uncertain and we estimate $c_1=0.36(10)$, see Appendix~\ref{aa:scal} (a quadratic dependence on $\log$ also gives an acceptable fit to both $\xi$ and $\mathcal{A}$ for the physical system, see also~\cite{Gorghetto:2020qws}). Therefore, at the physically relevant $\log(m_r/H_\star)\simeq 40\div 80$ for an ALP (for $m_a$ in the range $10^{-15}~\eV\div \eV$, c.f. Figure~\ref{fig:signal}), the area parameter is likely to be $\mathcal{A}_\star/N=\mathcal{O}(10\div100)$. Meanwhile, given the relation between $m_a$ and $f_a$, for the QCD axion $\log_\star\simeq 65$. As a result, a linear extrapolation predicts $\mathcal{A}_\star\simeq 10$ in this case (a quadratic extrapolation would instead give $\mathcal{A}_\star\simeq 100$).  Similarly to $\xi$, these values of $\mathcal{A}_\star$ are much larger than the $\mathcal{O}(1)$ values that occur at small scale separations.

In Figure~\ref{fig:arasim} (right) we show the effect of the $\mathbb{Z}_N$-preserving axion potential $V(a)$ in eq.~\eqref{eq:ax_pot} 
 with $\delta V_{PQ}=0$. In particular, we plot the evolution of $\mathcal{A}$ for the fat system for different choices of the axion mass $m_a$, which we parametrise by $\log(m_r/m_a)\equiv\log(m_r/H_\star)$, with time represented by $\log(m_a/H)$. To allow for a larger dynamical range,  $m_a$ is also taken to be proportional to $1/R$, so that the ratio $m_r/m_a$ is constant. (Although this is not physical, we expect the qualitative features of the dynamics to be similar, like in the fat string system.) For comparison, we also show the results obtained without any axion potential, $m_a=0$ (dashed lines, which correspond to the evolution in~Figure \ref{fig:arasim}, left). We plot the data starting from $\log(m_r/H)=3$, by which time strings are cleanly-defined objects. Note that the simulated values of $\log(m_r/m_a)$ are much smaller than the physically relevant ones. The results indicate that at $\log(m_a/H)\simeq 2$ the logarithmic growth of $\mathcal{A}$ stops. Not surprisingly, the increase in the area of the surfaces where $a/f_a=2\pi n$ with $\log_\star$ translates into a larger number of domain walls at this time (at least for the parameters possible in simulations) and the relation $\mathcal{A}\propto N$ remains approximately true.\footnote{As shown in Figure~\ref{fig:scalD} of Appendix~\ref{aa:scal}, the growth of $\xi$ continues unaffected slightly longer.} 

Figure~\ref{fig:arasim} (right) also shows a hint that $\mathcal{A}$ might decrease at large $\log(m_a/H)$ in the simulations with larger $\log_\star$. Although tentative, such a change is plausible and could indicate the domain walls, rather than the strings, starting to dominate the dynamics of the system (as might be expected once $\rho_{\rm w}\gtrsim \rho_{\rm s}$). Unfortunately, we do not have sufficient numerical reach to determine the asymptotic behaviour, e.g. it might be that the system is drawn to a new attractor with $\mathcal{A}$ independent of $\log_\star$ but we cannot determine whether this is the case.  
We also cannot rule out more dramatic qualitative changes in the network's dynamics at $\log_\star\gtrsim5$ and $\log(m_a/H) \gg 1$. Additionally, the non-linear transient of the waves produced during the scaling regime (expected to happen when $\log_\star\gg1$) could  affect the evolution of the domain walls around the time when $H=H_\star$ in a way that we do not have control of in both the case of an ALP and especially the QCD axion.

Given these multiple complicating factors, we have little control on the physically relevant value of $\mathcal{A}_d$, and thus leave  it unfixed in what follows. The resulting uncertainty on $f_a$ and $\tilde{g}_{a\gamma\gamma}$ is slightly tempered by the square-root in eq.~\eqref{eq:gaggD}.

\subsection{The emission spectrum}

As discussed above eq.~\eqref{eq:naw1}, the momentum of the axions emitted by the string-domain wall network is important for the axion relic abundance. The momentum distribution is encoded into the energy spectrum $\partial \rho_a/\partial k$, which can be extracted from numerical simulations as described in Appendix~\ref{app:sim}. However, while  domain walls are present they contaminate the axion field (and because they have thickness $\simeq m_a^{-1}$ they cannot be easily masked). Instead, it is more convenient to carry out simulations with non-zero $\mathbb{Z}_N$ breaking parameter, and extract the axion energy spectrum after all the domain walls are destroyed. However, we stress that we can only {determine the evolution and the spectrum} at very unphysical values of $m_a/m_r$ and $H_d/m_r$, and the true emission spectrum could differ from simulation results dramatically.

\begin{figure*}[t]
	\begin{center}
		\includegraphics[width=0.45\textwidth]{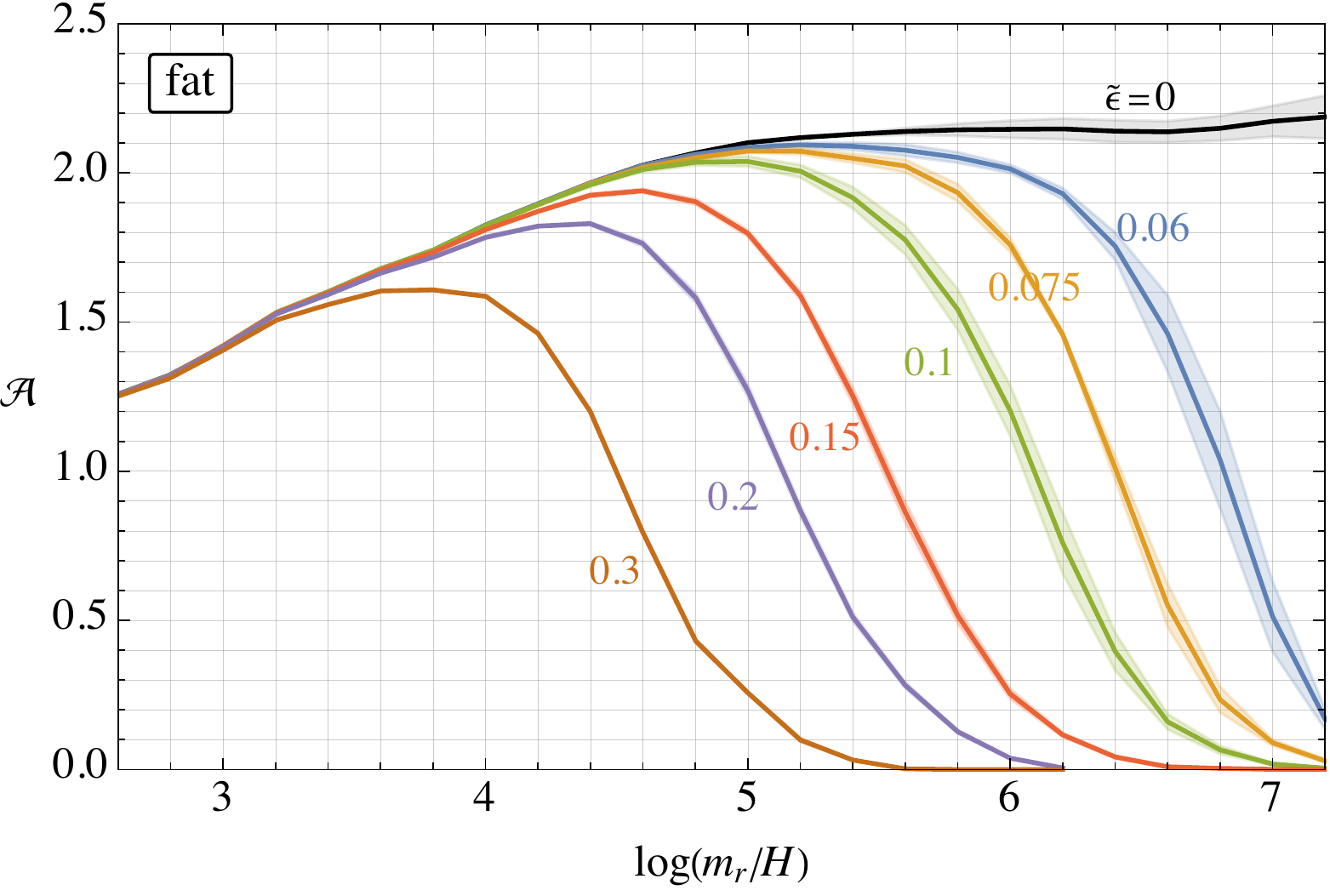} ~
		\includegraphics[width=0.525\textwidth]{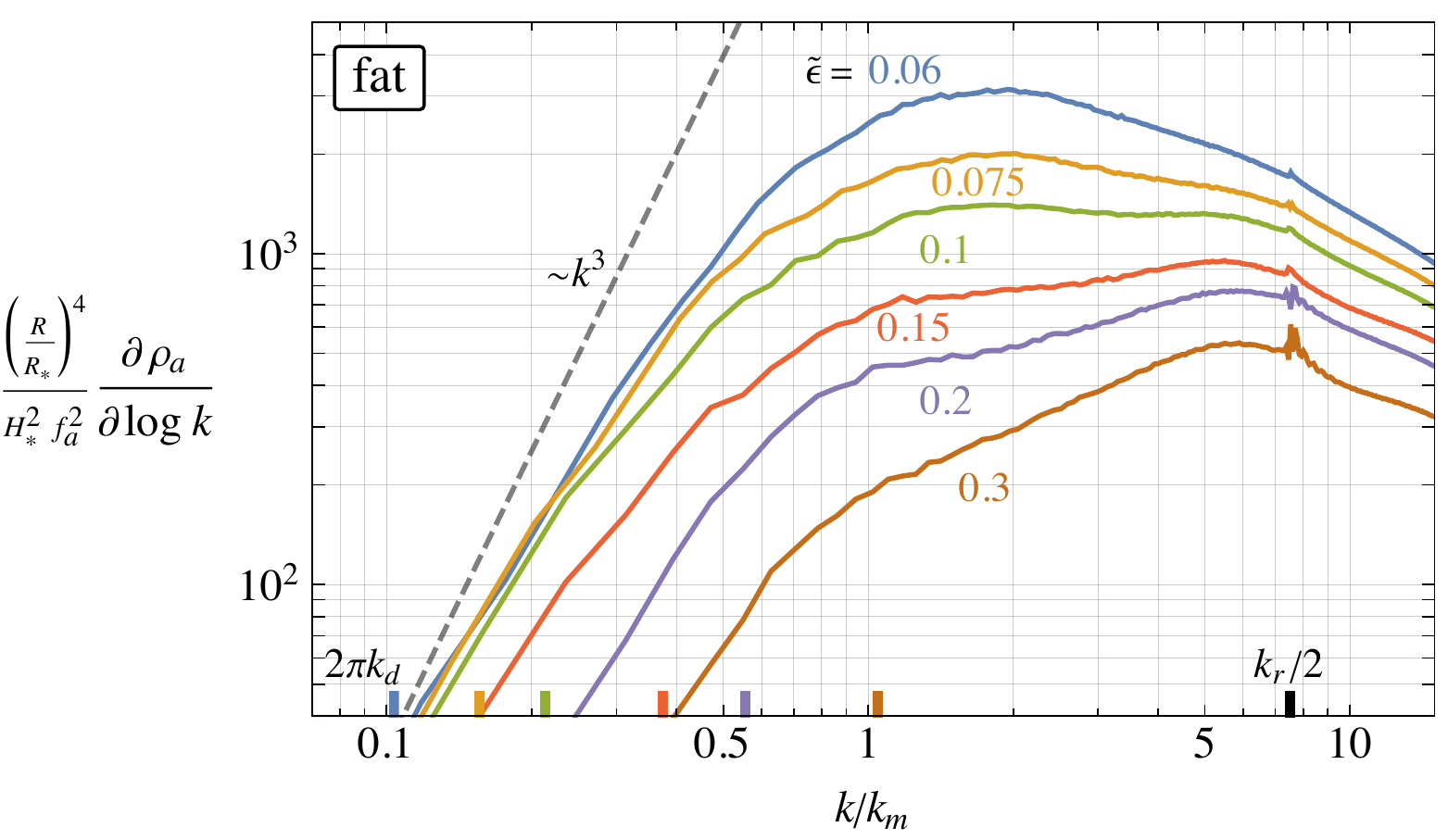}
	\end{center}
	\caption{ {\bf \emph{Left:}} The domain wall area parameter $\mathcal{A}$ in  fat string and domain wall theories with $N=4$ and $\log(m_r/m_a)=2.7$ for different values of the $\mathbb{Z}_N$ symmetry breaking parameter $\tilde{\epsilon}= v\sqrt{\epsilon}/(Nm_a)$.
		{\bf \emph{Right:}} The axion energy density spectra $\partial\rho_a/\partial \log k$  at the final simulation time, after the network is destroyed, for the simulations plotted in the left panel. The $x$-axis is normalised relative to the comoving momentum $k_m\equiv Rm_a$ corresponding to the axion mass $m_a$. We also indicate the comoving momentum corresponding to the string core scale $k_r/2$. The coloured ticks indicate $2\pi k_d$, where $k_d$ is the comoving momentum that is equal to $H$ at the time when the network is destroyed. For sufficiently small $\tilde{\epsilon}$ (i.e. a relatively long lived network), the axions seem to be emitted by domain walls at most mildly relativistic, with a spectrum peaked at momenta of the order of the axion mass $k\simeq k_m$, but it is challenging to extrapolate this result at the physical parameter space. 
		\label{fig:scalD}}
\end{figure*}

In Figure~\ref{fig:scalD} (left) we plot $\mathcal{A}$ as a function of time in such simulations for the fat string-domain wall system 
with different amounts of $\mathbb{Z}_N$ breaking, which we parameterise by $\tilde{\epsilon}= v\sqrt{\epsilon}/(Nm_a)$ where $\epsilon$ is defined in eq.~\eqref{eq:ax_pot2}. We fix $\log(m_r/H_\star)=2.7$ (i.e. $m_r/m_a=15$) and $N=4$. As expected, larger $\tilde{\epsilon}$ results in the network being destroyed earlier.\footnote{We do not attempt to fit a relation between the $\tilde{\epsilon}$ and the Hubble parameter when the network is destroyed $H_d$, which would require extremely careful extrapolations in $m_r/m_a$ and $m_a/H_d$.} 
In Figure~\ref{fig:scalD} (right) we plot the axion energy spectrum at the end of the simulations shown in the left panel, by which time all the strings and domain walls are destroyed, the axion field amplitude is $\ll f_a$ over all of space (apart from a few oscillons~\cite{Vaquero:2018tib}, which contain a negligible fraction of the axion energy density), and the spectrum has reached a fixed shape. 
Because of the fat strings and domain walls, axions produced with proper momentum $m_a$ or $m_r/2$ (the latter corresponding to the string core scale) always have, and remain at, the same comoving momentum $k_m=m_a(t) R(t)$ or $k_r/2= m_r(t) R(t)/2$ respectively. At the values of $\log(m_r/H)$ that can be simulated, the strings are expected to emit energy dominantly to axions with UV momentum $m_r$~\cite{Gorghetto:2018myk,Gorghetto:2020qws} although this has only been shown with $m_a=0$.

In Figure~\ref{fig:scalD} (right), the axion energy spectrum increases as $\tilde{\epsilon}$, and therefore $H_d$, decreases. This indicates that, at least for the smallest $\tilde{\epsilon}$, a sizable fraction of the energy density in axion waves is coming from the domain walls (because the relic abundance from strings is dominantly produced at $H=H_\star$ rather than later). Although $m_r/m_a=15$ only gives a mild separation between $k_a$ and $k_r/2$, Figure~\ref{fig:scalD} suggests that with longer lived domain walls an increased fraction of the axion waves are produced at most borderline rather that ultra-relativistic, with physical momenta $\simeq m_a$ rather than $m_r/2$ (or equivalently, comoving momentum $\simeq k_m$ rather than $k_r/2$). Although at these scale separations the emission spectrum from domain walls is peaked at $k_m$ rather than $k_d$ (the latter of which is set by the Hubble parameter when the networks is destroyed),\footnote{We have checked that results from simulations with different $N$ and $m_r/m_a$ are consistent.} 
 we cannot exclude that this changes at much larger values of $\log(m_r/m_a)$ and $\log(m_a/H)$, and we leave a full extrapolation for future work.

\section{Observational implications} \label{sec:signals}

\subsection{Constraints} \label{ss:constraints}

There are numerous constraints on $g_{a\gamma\gamma}$ that apply to either any axion or any axion that is DM. Additionally there are bounds that are specific to post-inflationary axions. 

Given that their energy redshifts  differently  to dark matter (in particular, slower), the domain walls must decay before matter-radiation equality (MRE) if the axion makes up a sizable component of dark matter. In the $N>1$ case, this requires that the Hubble parameter  when the network is destroyed $H_d$ is larger than $H_{\rm eq}$, where `eq' denotes quantities at MRE (similarly, in $N=1$ theories $H_\star>H_{\rm eq}$ is needed). If the axion makes the whole DM, using eq.~\eqref{eq:relic}, this constrains
\beq \label{eq:Hdeq}
f_a \gtrsim  10^7\GeV \frac{1}{\mathcal{A}_d^{1/2}}  \left(\frac{10^{-6}~\eV}{m_a} \right)^{1/2}
~.
\eeq

Observational limits on the fraction of primordial isocurvature density perturbations 
lead to less certain, but probably stronger, constraints on both $N=1$ \cite{Feix:2019lpo,Irsic:2019iff,Feix:2020txt,Gorghetto:2021fsn} and $N>1$ post-inflationary theories. 
These require any primordial isocurvature fluctuations (such as the axion's ones), with power spectrum $\mathcal{P}_{\rm iso}(k)$, to be suppressed on observable scales 
with respect to the initially small adiabatic fluctuations, described by the almost scale-invariant power spectrum of curvature perturbations, $\mathcal{P}_{\mathcal{R}}(k)=A_s(k/k_p)^{n_s-1}$, where $k_p/R_0=0.05\,{\rm Mpc}^{-1}$ is a pivot scale and $A_s=2.2\times 10^{-9}$, with $n_s\simeq1$. 
In particular, at masses $m\gtrsim 10^{-22} \,\eV$, the strongest such bound comes from Lyman-$\alpha$ observations (looking at momenta $k/R_0=k_{L\alpha}/R_0\equiv 10\, {\rm Mpc}^{-1}$) and limits the fraction of isocurvature perturbations as $f_{\rm iso}^2(k)\equiv \mathcal{P}_{\rm iso}(k)/\mathcal{P}_{\mathcal{R}}(k)\lesssim 0.004$ at $k=k_p$ assuming $\mathcal{P}_{\rm iso}(k) \propto k^3$ for $k\in (k_p,k_{L\alpha})$~\cite{Irsic:2019iff}. 

To have a chance of being compatible with isocurvature constraints, the axion modes on this scale must be in the $k^3$ part of the power spectrum, so the resulting constraint should only depend on the numerical coefficient $C$ and $m_a$. For $N=1$, using eq.~\eqref{eq:Pdelta}, $f^2_{\rm iso}=(\Omega_a/\Omega_{\rm DM})^2(C k_p/k_\star)^3/A_s$ and therefore this isocurvature bound corresponds to
\beq \label{eq:isoN1}
m_a \gtrsim 2\cdot 10^{-17} C^2 \left(\frac{\Omega_a}{\Omega_{\rm DM}} \right)^{4/3}\,\eV ~.
\eeq
where, based on numerical simulations at small scale separation, a value of order $C =\mathcal{O}(10^{-1})$ is plausible~\cite{Gorghetto:2021fsn}.

Also for $N>1$ the constrained scales must lie in the $k^3$ part of $\mathcal{P}_\delta$.  
If the network decays at $H_d \ll m_a$ and turns out to indeed have a peak at $k\simeq k_d$, the resulting bound is stronger than in the $N=1$ case because the order one perturbations are on much larger spatial scales $k_d\ll k_\star$, and $\mathcal{P}_\delta$ at observable scales is less suppressed by the $k^3$ form. Under this {reasonable} assumption, $H_d$ plays exactly the role that $m_a$ does in the $N=1$ case, so the constraint in \cite{Irsic:2019iff} can be simply translated to
\beq \label{eq:isoN}
H_d \gtrsim 2\cdot 10^{-17} C^2 \left(\frac{\Omega_a}{\Omega_{\rm DM}} \right)^{4/3}\,\eV  ~.
\eeq
Combining eq.~\eqref{eq:isoN} with the result for the relic abundance eq.~\eqref{eq:relic} gives the bound from isocurvature perturbations
\beq \label{eq:faIso}
f_a\gtrsim 6\times10^{9}~\GeV \frac{1}{\mathcal{A}_d^{1/2}}  \left(\frac{10^{-6}~\eV}{m_a} \right)^{1/2} C^{1/2} \left(\frac{\Omega_a}{\Omega_{\rm DM}} \right)^{5/6}~,
\eeq
and a corresponding upper bound on $\tilde{g}_{a\gamma\gamma}$. 
As discussed, we cannot reliably extract the value of $C$ in the $N>1$ scenario from numerical simulations. However, based on results for the $N=1$ string-wall network, a reasonable expectation is $C=\mathcal{O}(10^{-1}\div1)$, but not too much smaller~\cite{Gorghetto:2021fsn}. Note that the bound on $f_a$ (and therefore $\tilde{g}_{a\gamma\gamma}$) is weaker for larger $\mathcal{A}_d$ because this provides the correct dark matter abundance for larger $H_d$.

If the axions are produced with initial momentum of order $m_a$, there is a constraint from the axions behaving as warm dark matter because the free-streaming of the dark matter particles would suppress the formation of small-scale structures \cite{Bode:2000gq,Viel:2013fqw,Gelmini:2022nim}. In theories of thermally produced warm dark matter,  the observational bound on the dark matter mass is $m_{\rm WDM}\gtrsim 10 \keV$, i.e. the dark matter must be at least borderline non-relativistic when the Universe has temperature $T=10 \keV$ \cite{Irsic:2017ixq,Baumholzer:2020hvx,Dekker:2021scf}. The free streaming length, and therefore the resulting cut-off in the linear matter power spectrum, only depends on the dark matter equation of state \cite{Ballesteros:2020adh}. Therefore, a limit on $N>1$ theories can be obtained by demanding that, if the axions are produced semi-relativistic, $H_d \gtrsim H_{T=10 \keV}$ where $H_{T=10 \keV}$ is the value of the Hubble parameter when $T=10 \keV$. This leads to the constraint on an axion that makes up all the dark matter of~\footnote{There is uncertainty on this bound both from the unknown details of the initial power spectrum, e.g. how relativistic the axions are at $H_d$ and also from the fact we did not do a full analysis of the effects of warm, low mass, axions and instead reinterpreted existing bounds. With a reliable prediction of the energy spectrum of the axions produced by the domain walls, one could obtain an accurate constraint by following the approach of \cite{Ballesteros:2020adh}.}
\beq \label{eq:warm_bound1}
f_a\gtrsim 10^{9}~\GeV \frac{1}{\mathcal{A}_d^{1/2}}  \left(\frac{10^{-6}~\eV}{m_a} \right)^{1/2} ~,
\eeq
and a corresponding one on $\tilde{g}_{a\gamma\gamma}$. The resulting bound is approximately an order of magnitude weaker than the isocurvature one with $C\simeq \mathcal{O}(0.1)$. 
If $\Omega_a<\Omega_{\rm DM}$ the warm dark matter bound is less certain. Fitting the numerical results of \cite{Diamanti:2017xfo} suggests that, approximately, the minimum temperature at which the axions must be non-relativistic (which coincides with the destruction temperature $T_d$ with our present assumptions) is $T_{d} \propto \left(\Omega_{a}/\Omega_{\rm DM} \right)^{1.5}$  in which case the allowed $f_a$ in eq.~\eqref{eq:warm_bound1} would scale as $ \propto \left(\Omega_{a}/\Omega_{\rm DM} \right)^{1.25}$. (Note that, as shown in Ref.~\cite{Amin:2022nlh}, the combination of the isocurvature bound and the constraint on warm dark matter leads to the limit $m_a\gtrsim 10^{-18}~\eV$ in post-inflationary theories regardless of the scale $k$ at which $\mathcal{P}_\delta$ is of order one.)

We note that the explicit $\mathbb{Z}_N$ symmetry breaking parameter $\epsilon$ enters in all of the constraints only via $H_d$, so the uncertainty on $H_d$ as a function of the $\epsilon$ does not propagate to eqs.~\eqref{eq:Hdeq}, \eqref{eq:faIso}, and \eqref{eq:warm_bound1}.

\begin{figure*}[t!]
	\begin{center}
		\includegraphics[width=0.95\textwidth]{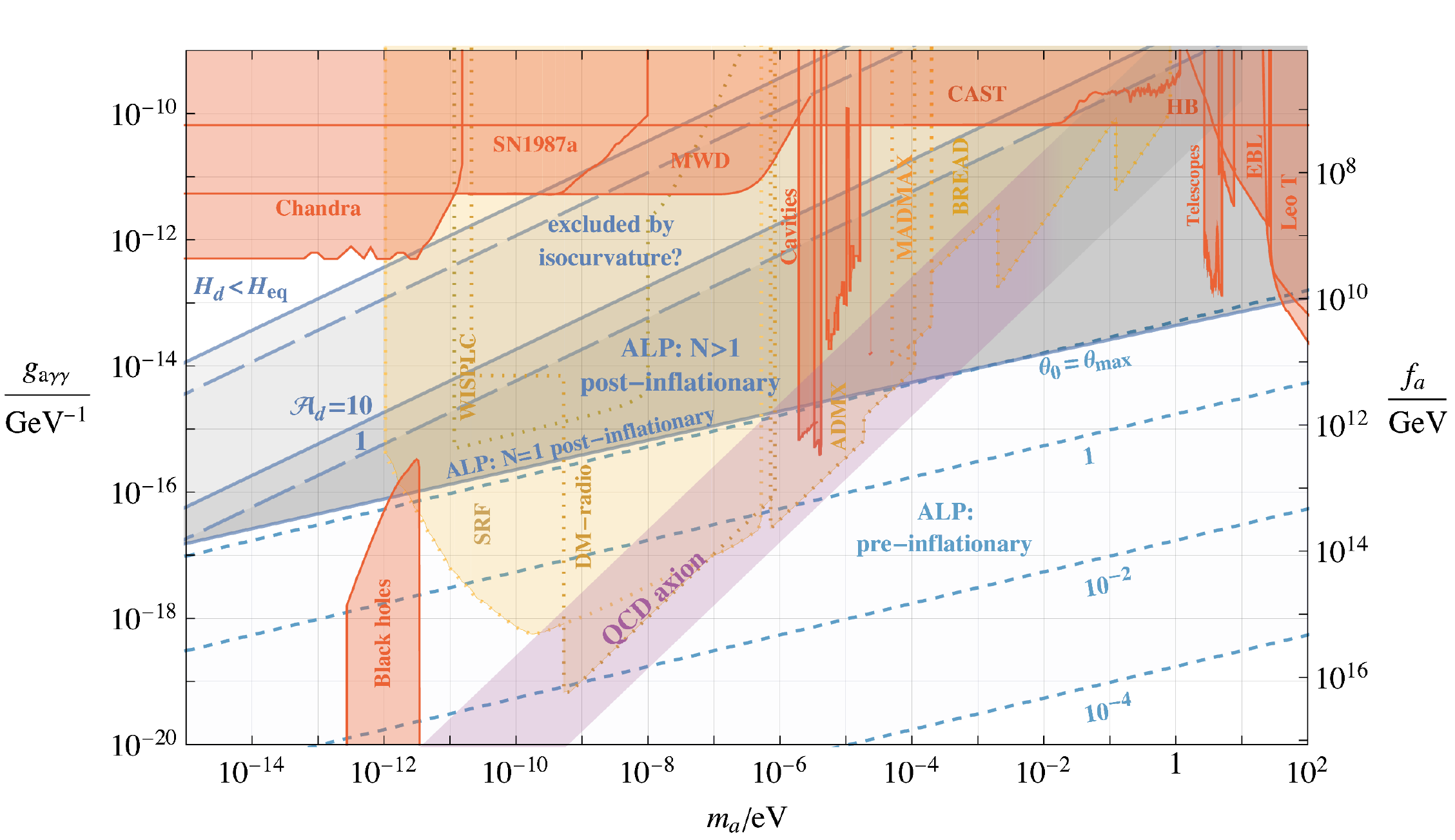}
	\end{center}\vspace{-1mm}
	\caption{ 
       	\emph{Blue colours:}   The axion-to-photon coupling $g_{a\gamma\gamma}$ as a function of the axion mass $m_a$ in different minimal scenarios, requiring that an ALP makes up the full DM relic abundance. We assume that this coupling is given in terms of $f_a$ as $g_{a\gamma\gamma}\equiv \alpha_{em}/(2\pi f_a)$, i.e. $E/N=1$ in eq.~\eqref{eq:gagg}; in a theory with $E/N\neq 1$ the corresponding $g_{a\gamma\gamma}$ will be shifted. Dashed lines represent the pre-inflationary scenario for different misalignment angles, $\theta_0$. $\theta_{\rm max}$ is the largest such angle compatible with isocurvature constraints if the Hubble scale during inflation $H_I=1\eV$ (smaller $\theta_0$ are required if $H_I$ is larger). 
       The blue shaded region represents the  post-inflationary scenario with $N>1$ as the parameter $\epsilon$, which controls the $\mathbb{Z}_N$-breaking potential, varies. The area labelled `excluded by isocurvature' is likely to be in tension with isocurvature constraints, although these depend on the axion's poorly known density power spectrum after the network decays; the results are plotted assuming the coefficient of the white-noise tail in eq.~\eqref{eq:Pdelta} to be $C=0.1$. The region `$H_d<H_{\rm eq}$' is  excluded because the domain walls would be destroyed after matter-radiation equality. Both these constraints depend on the  domain wall area parameter at the time of the network's decay $\mathcal{A}_d$, which is uncertain (we plot the results for $\mathcal{A}_d=1$ and $10$, dashed and solid). The lower boundary of the $N>1$ region very approximately coincides with the $N=1$ scenario. 
		\emph{Red/orange colours:} Existing experimental and observational bounds and the reach of some proposed future haloscopes (orange)  \cite{AxionLimits}.
		\label{fig:signal}}
\end{figure*}

\subsection{Relevance for experimental searches} \label{ss:experiment}
The ALP-to-photon couplings in the pre- and post-inflationary scenarios are shown in Figure~\ref{fig:signal} imposing that $\Omega_a=\Omega_{\rm DM}$. To enable comparison with existing and projected constraints we assume $E/N=1$, so $g_{a\gamma\gamma}=\tilde{g}_{a\gamma\gamma}\equiv\alpha_{em}/(2\pi f_a)$.\footnote{The current constraints on  $g_{a\gamma\gamma}$ in Figure~\ref{fig:signal} come from various sources \cite{Reynes:2021bpe,Dessert:2020lil,Meyer:2020vzy,Dessert:2022yqq,CAST:2017uph,Cadamuro:2011fd,Wadekar:2021qae,Regis:2020fhw,Grin:2006aw}  (we do not attempt to be exhaustive and simply present the most important ones in each mass range). We also plot the bounds from black hole super-radiance, which do not require an axion-photon coupling \cite{Arvanitaki:2014wva,Baryakhtar:2020gao} and instead are relevant only for $f_{a}$ not too small, because a too large axion self-interaction hinders the super-radiance process.} 
The axion-photon coupling for the QCD axion, given in eq.~\eqref{eq:ggamma_qcd}, is also shown; we set the upper and lower edges of the band by $E/N=44/3$ and $2$ respectively, but we stress that couplings outside the range plotted are possible (the band is grayed for $m_a \gtrsim 10^{-2}~\eV$ because astrophysical  observations rule out many such QCD axion theories   \cite{Raffelt:2006cw,Chang:2018rso,Carenza:2019pxu,Viaux:2013lha,MillerBertolami:2014rka}).

In Figure~\ref{fig:signal}, dashed lines show the values of the initial misalignment angle $\theta_0$ for which an ALP in the pre-inflationary scenario makes up the full relic density. As can be seen, such an axion with $E/N\simeq 1$ can be all DM and compatible with existing constraints for almost any $m_a\in (10^{-22}~\eV ,10^2~\eV)$ \cite{Irsic:2017yje,Lora:2011yc,Gonzalez-Morales:2016yaf,Marsh:2018zyw,Cadamuro:2011fd}. 
Meanwhile, an ALP in the post-inflationary $N>1$ scenario can make up the full dark matter abundance over the dark blue area as the value of the explicit symmetry breaking parameter $\epsilon$, which determines $H_d$,  varies. Points with smaller $f_a$ (i.e. larger $\tilde{g}_{a\gamma\gamma}$) correspond to smaller $\epsilon$, i.e. smaller $H_d$. The upper edge of this region is set by the lowest $H_d$ allowed by isocurvature constraints, which depend on the uncertain parameters $\mathcal{A}_d$ and $C$ (we also plot the weaker bound from requiring the domain wall network is destroyed before MRE). 
The lower edge of the $N>1$ post-inflationary region corresponds to $\epsilon\simeq m_a^2/f_a^2 $ so  $H_d\simeq m_a$ and the domain walls are destroyed soon after forming. In this case the DM abundance, and therefore $\tilde{g}_{a\gamma\gamma}$, is of the same order as that in the $N=1$ post-inflationary  scenario, which we calculate considering only the  abundance produced by strings eq.~\eqref{eq:string_relic}  (if domain walls give a large contribution to the relic abundance the prediction would be at larger $g_{a\gamma\gamma}$).

\subsection{Possible implications for gravitational wave searches}

In a post-inflationary $N>1$ axion theory the evolution of the domain wall network between $H_\star$ and $H_d$ produces a stochastic GW background \cite{Chang:1998tb,Hiramatsu:2012sc,Gelmini:2021yzu,Gelmini:2022nim}. Due to the uncertainties in the dynamics of the network, the rate of production of these GWs is unknown. However, standard assumptions making use of the  quadrupole approximation predict that the domain walls emit energy density in GWs at a rate $\Gamma_{\rm GW}$ roughly given by (see e.g. \cite{Maggiore:2007ulw,Hiramatsu:2012sc,Gelmini:2021yzu})
\beq
\begin{aligned} \label{eq:GGW}
\Gamma_{\rm GW}/H & \simeq \mathcal{A}^{2} G \sigma^2 \\
& \simeq \mathcal{A}^{2} G m_a^2 f_a^4 ~.
\end{aligned}
\eeq
If eq.~\eqref{eq:GGW} is accurate, the energy lost into GWs is much smaller than the total energy the network emits (in eq.~\eqref{eq:GammaW}) over all of the allowed parameter space in Figure~\ref{fig:signal}, and their backreaction is therefore not expected to affect the dynamics of the domain walls.

Because the domain walls have typical curvature of order Hubble, the instantaneous emission spectrum of the GWs is expected to be peaked at the Hubble parameter at the time of production. This is indeed the case in numerical simulations at small scale separations \cite{Hiramatsu:2012sc}. If true with the physical parameters, the  GW spectrum today $\Omega_{\rm GW}(f)\equiv \frac{1}{\rho_{\rm c}} \partial\rho_{\rm GW}/\partial\log f$, where $f$ is the present-day GW frequency, will be peaked at the frequency corresponding to the Hubble parameter at the time the network is destroyed $	f^{\rm peak}  \simeq  10^{-16}~{\rm Hz} \left(H_d/H_{\rm eq} \right)^{1/2}$, i.e.
\beq \label{eq.fGW1}
	f^{\rm peak} \simeq 7\cdot 10^{-7} ~{\rm Hz}~ \mathcal{A}_d  \left(\frac{\Omega_a}{\Omega_{\rm DM}} \right)^{-1} \left(\frac{f_a}{10^{12}~\GeV} \right)^2 \left(\frac{m_a}{10^{-6}~\eV} \right) ~.
\eeq
Assuming that eq.~\eqref{eq:GGW} reproduces the actual emission, the amplitude of the GW spectrum at its peak is
\beq \label{eq.OGW1}
\begin{aligned}
	\Omega_{\rm GW}|_{\rm peak} &\simeq 10^{-31} \mathcal{A}_d^{2}  \left(\frac{f_a}{10^{12}~\GeV} \right)^4  \left(\frac{ m_a}{H_d}\right)^2
	\\
	&\simeq 5\cdot 10^{-28} \frac{1}{\mathcal{A}_d^{2}} \left(\frac{\Omega_a}{\Omega_{\rm DM}} \right)^4   \left(\frac{ 10^{12}~\GeV}{f_a} \right)^{4} \left(\frac{10^{-6}~\eV}{m_a} \right)^2 .
\end{aligned}
\eeq
Eq.~\eqref{eq.fGW1} and the second line of eq.~\eqref{eq.OGW1} follow from  the expression for the axion relic abundance in eq.~\eqref{eq:relic} and we allow $\Omega_a\neq \Omega_{\rm DM}$ for generality. As is evident from eq.~\eqref{eq.OGW1}, the strongest GW signals occur for the smallest allowed $f_a$. This is because, for fixed $m_a$ and $\Omega_a$, smaller $f_a$ corresponds to smaller $H_d$, so the GWs are redshifted less. 

The possibility of observing these GWs is in tension with the observational bound on isocurvature perturbations described in Section~\ref{ss:constraints}: combing  eq.~\eqref{eq.OGW1} with  eq.~\eqref{eq:faIso}, the isocurvature bound implies
\beq \label{eq:OGWmax}
\Omega_{\rm GW}|_{\rm peak} \lesssim 4\cdot 10^{-19} \frac{1}{C^{2}} \left(\frac{\Omega_a}{\Omega_{\rm DM}} \right)^{2/3}~,
\eeq
independent of $m_a$ and $\mathcal{A}_d$. For $C\gtrsim \mathcal{O}(10^{-2})$ we have $\Omega_{\rm GW}|_{\rm peak} \lesssim 10^{-15}$, and {if the axion is only a small fraction of DM, this limit is not relaxed.}\footnote{Meanwhile, the warm dark matter bound eq.~\eqref{eq:warm_bound1}, which applies if the axions are produced mildly relativistic,  leads to $\Omega_{\rm GW}|_{\rm peak} \lesssim 5\cdot 10^{-16} \mathcal{A}_d^{-1} \left(\Omega_a /\Omega_{\rm DM} \right)^{-1} $. This is weaker than eq.~\eqref{eq:OGWmax} for $C\gtrsim 10^{-2}$ (see also \cite{Gelmini:2022nim} for a related analysis).} Near-future detectors are likely to only reach sensitivities $\Omega_{\rm GW} \simeq 10^{-15}$ for any frequency predicted by eq.~\eqref{eq.fGW1}.\footnote{Pulsar timing arrays will, at best, reach $\Omega_{\rm GW} \simeq 10^{-15}$ for frequencies around $\simeq 10^{-8}~{\rm Hz}$, and space based interferometers could have similar reach at higher frequencies. Because GWs redshift like radiation, the energy density they contain can be constrained by CMB observations \cite{Smith:2006nka}, but current bounds are in the region of  $\Omega_{\rm GW} \simeq 10^{-6}$ \cite{Pagano:2015hma} and could be improved by only a few orders of magnitude (e.g. by the future EUCLID satellite). Finally, at frequencies $\lesssim 10^{-14}~{\rm Hz}$ measurements of the CMB polarisation data are relevant (see e.g. \cite{Kamionkowski:1999qc}). Current limits are $\Omega_{\rm GW} \simeq 10^{-12}$ for GWs of frequency $f_{\rm eq}\simeq 10^{-16}~\rm Hz$ \cite{Planck:2018vyg,BICEP2:2018kqh} (corresponding to $H_d = H_{\rm eq}$) and  at higher frequencies these get weaker $\propto (f_{\rm eq}/f)^2$ (planned experiments e.g. LiteBIRD \cite{Matsumura:2013aja}, could have  a factor $\sim 100$ better sensitivity).} Therefore, unless the unknown constant $C$ (which parametrises the white-noise tail of the axion density power spectrum, see eq.~\eqref{eq:Pdelta}) is much smaller than the naive expectation, the allowed GW signals from ALP domain walls are barely detectable regardless of $m_a$ and $f_a$. We do however stress that  eq.~\eqref{eq:GGW} is just a rough parametric expectation, and an analysis of the GW emission from domain walls in numerical simulations taking into account the several required extrapolations would certainly be worthwhile, and left for future work. We also note that it is possible  that the network stops emitting energy to axions efficiently enough to maintain an approximately constant $\mathcal{A}$ at scale separations beyond the reach of simulations, completely altering the network's dynamics and the resulting GW prediction.

GWs are also produced by the axion strings prior to $H_\star$ \cite{Gorghetto:2021fsn} and the resulting stochastic background is detectable, compatibly with constraints from isocurvature perturbations, dark radiation, and axion dark matter overproduction, if  $f_a \gtrsim 10^{14}~\GeV$, and $10^{-20} ~\eV< m_a< 10^{-18}~\eV$. For such high $f_a$ however typical $g_{a\gamma\gamma}$ are suppressed and the relic axions cannot be directly detected (except for specific models e.g. based on \cite{Choi:2015fiu,Kaplan:2015fuy}, where $g_{a\gamma\gamma}\ggg1/f_a$).

\subsection{Comments on ALP dark matter substructure}

As discussed in Section~\ref{ss:power}, the decay of the string-domain wall network leads to order one fluctuations in the axion dark matter overdensity field at small spatial scales, which might lead to interesting dark matter substructure \cite{Hogan:1988mp,Kolb:1993zz,Zurek:2006sy,Hardy:2016mns,OHare:2021zrq}, see \cite{Fairbairn:2017sil,Visinelli:2018wza,Eggemeier:2019khm,Ellis:2020gtq,Ellis:2022grh,Dandoy:2022prp,Shen:2022ltx} for analysis in the case of the QCD axion.  Although their properties are extremely uncertain, we now briefly comment on the evolution of an ALP's initial fluctuations assuming that there are order-one overdensities on scales $H_d^{-1}$. As we will see, quantum pressure and axion self-interactions can play an important, previously neglected, role.

After the string-wall decays, the axion follows the nonrelativistic Schr\"odinger--Poisson EoM, which can be rewritten as the continuity and Euler equation of the density and velocity fields, $\rho$ and $v$. At the small spatial scales we are interested in, $\delta\equiv (\rho_a-\bar{\rho}_a)/\bar{\rho}_a=\mathcal{O}(1)$. Nevertheless, we can obtain a qualitative understanding of the evolution of a generic mode (with comoving momentum $k$) by considering the linearised form of the EoM 
\begin{equation}\label{eq:delta_pert_1}
\ddot{\delta}+2H\dot{\delta}-\left[4\pi G\rho_a-\frac{k^2}{R^2}\left(\frac{k^2}{4m^2R^2}+\frac{g\rho_a}{m_a^2}\right)\right]\delta=0 \, ,
\end{equation}
where $g\equiv\lambda/m_a^2\simeq- 1/(8f_a^2)$ parametrises the  axion quartic coupling $V\supset \lambda a^4/3$, which we assume is attractive as is the case for a cosine potential (higher order couplings are present but negligible).  The square bracket in eq.~\eqref{eq:delta_pert_1} is the Laplacian of the total potential $\Phi+\Phi_Q+\Phi_{\rm self}$ acting on the field, where $\Phi_Q\equiv- \left(\nabla^2\sqrt{\rho_a}\right)/(2m_a^2R^2\sqrt{\rho_a})$ and $\Phi_{\rm self}\equiv (g\rho_a)/m_a^2$ are the `quantum' pressure and self-interaction potentials respectively, while $\Phi$ is the gravitational one, satisfying $\nabla^2\Phi=4\pi GR^2\rho_a$. Both $\Phi$ and $\Phi_{\rm self}$  are attractive and tend to increase $\delta$ (given that $g<0$), while quantum pressure is repulsive and makes $\delta$ oscillate, preventing perturbations from collapsing. Note that eq.~\eqref{eq:delta_pert_1} reduces to the usual equation describing cold dark matter perturbations (the Meszaros equation) in the limits $m\to\infty$ and $g\to0$. We refer to~\cite{Arvanitaki:2019rax,Chavanis:2016dab,Chavanis:2011uv,Hwang:2009js,Suarez:2015fga} for the derivation of eq.~\eqref{eq:delta_pert_1}.

The EoM~\eqref{eq:delta_pert_1} takes a simplified form when expressed in terms of $y\equiv R/R_{\rm eq}$,
\begin{equation}\label{eq:delta_pert_2}
\delta''+\frac{3 y+2}{2 y (y+1)}\delta'-\frac{3}{2 y (y+1)}\left[1-\left(\frac{k}{ k_{J}^{\rm eq}}\right)^4\frac{1}{y}+\left(\frac{k}{ k_{\rm self}^{\rm eq}}\right)^2\frac{1}{y^2}\right]\delta =0 \, ,
\end{equation}
where $\delta'\equiv d\delta/dy$ and $k_J$ and $k_{\rm self}$ are two comoving momentum scales associated with quantum pressure (`quantum' Jeans scale~\cite{Hui:2021tkt}) and self-interactions, defined by 
\beq \label{eq:kJ}
k_J\equiv R (16\pi G\rho_a m_a^2 )^{1/4}~, \qquad\qquad k_{\rm self}\equiv R m_a(4\pi G/|g|)^{1/2}~,
\eeq
respectively, and the superscript `eq' refers to quantities calculated at $R=R_{\rm eq}$.  
For very IR modes ($k/k_{J}^{\rm eq}\ll1$ and $k/k_{\rm self}^{\rm eq}\ll1$, which includes the normal adiabatic modes at cosmological scales) quantum pressure and self-interactions are negligible, and eq.~\eqref{eq:delta_pert_2} shows that such perturbations increase linearly around MRE as $\delta\propto 1+(3/2)R/R_{\rm eq}$ while order-one fluctuations collapse into bound objects at $R=R_{\rm eq}/\delta$. However, for large enough $k$ quantum pressure and self-interactions affect the evolution by preventing the modes from collapsing, or enhancing their growth.

Interestingly, at MRE the quantum Jeans scale $k_J^{\rm eq}$ coincides with $k_\star$, modulo an order one factor: 
\beq \label{eq:coin}
\frac{k_\star}{k_J^{\rm eq}}
\simeq 0.53~,
\eeq
similarly to vector bosons produced by inflationary fluctuations~\cite{Gorghetto:2022sue}. This relation holds regardless of the value of $m_a$ and even Newton's constant (but it is modified if the axion mass is temperature dependent).\footnote{In eq.~\eqref{eq:coin} we assume that the number of relativistic degrees of freedom $g_\star$ at $H_\star$ is given by the standard model high temperature value, which is appropriate for $m_a\gtrsim 10^{-5}\eV$. However, the numerical factor in eq.~\eqref{eq:coin} only depends on $g_\star^{1/12}$, so this has a negligible impact on our conclusions (e.g. of order 100 beyond Standard Model degrees of freedom around the TeV scale would not affect our conclusions).}  

\subsubsection*{$\pmb{N=1}$}

In theories with $N=1$ the density power spectrum $\mathcal{P}_\delta$ is expected to be peaked at 
   $k_{\rm peak}= x k_\star$ with $\mathcal{P}_\delta(k_{\rm peak})\sim \mathcal{O}(1)$, where $x$ is a numerical factor that is highly uncertain (simulations at $\log(m_r/H_\star)\simeq 5$ suggest $x\sim \mathcal{O}(10)$ \cite{Gorghetto:2021fsn} although this could be modified prior to MRE by free-streaming or, as we discuss shortly, self-interactions). Therefore, given eq.~\eqref{eq:coin}, the collapse of order-one fluctuations will be affected by quantum pressure unless $x\ll1$, which would be surprising.   In particular if $x \gtrsim 2$ quantum pressure will prevent fluctuations collapsing at MRE, and compact objects are not expected to form from these fluctuations.

Meanwhile, for all the values of $m_a$ and $f_a$ of interest 
$
k_\star/k_{\rm self}^{\rm eq}\ll1 
$,
so at MRE the self-interactions do not  affect the modes at $k\simeq k_\star$ or smaller. 
However, self-interactions could be important deep in radiation domination,\footnote{We thank Asimina Arvanitaki for pointing this out.} when the corresponding terms in eqs.~\eqref{eq:delta_pert_1} and eq.~\eqref{eq:delta_pert_2} can be larger than the quantum pressure term because $\Phi_g$ is proportional to the density. 
By solving eq.~\eqref{eq:delta_pert_2} starting at $R_0=R_\star$ with $\delta'(R_0)=0$, one can see that indeed perturbative modes at momenta $k\gtrsim 10^{-2} k_\star$ grow substantially during radiation domination due to the self-interactions. 
This could significantly modify the initial spectrum of density fluctuations before the overdensities collapse. 
Given the large uncertainties on the initial spectrum, we leave a full study for future work.

\subsubsection*{$\pmb{N>1}$}
Conversely, for $N>1$ the string network decays at a time when $H_d<H_\star$ and the order one fluctuations are expected to lie on larger comoving scales, $x k_d$.  
The numerical value of $x$, although unknown, might be $\mathcal{O}(10)$, similarly to the $N=1$ scenario. At MRE, the ratio between the size of these fluctuations and the quantum Jeans scale reads 
\beq \label{eq:kdkJ}
\begin{aligned}
\frac{k_{\rm peak}}{k_J^{\rm eq}} & \simeq  x \left( \frac{H_d}{m_a} \right)^{1/2} ~.
\end{aligned}
\eeq
Thus, for $H_d\ll m_a$ (i.e. well within the blue region in the parameter space in Figure~\ref{fig:signal}), quantum pressure is not expected to be relevant in the dynamics around MRE. Additionally, by solving eq.~\eqref{eq:delta_pert_2} starting at $R_0=R_d$, self-interactions are also negligible throughout for modes $k\simeq k_d$ or smaller. 
Thus, the order-one perturbations are expected to collapse into compact objects (miniclusters) with density $\sim \bar{\rho}(R_{\rm eq})$,  
i.e. set by the average dark matter density when they collapse. The mass of the miniclusters will be similar to the mass contained within an initially order one fluctuation
\beq
\begin{aligned} \label{eq:M0}
	M_d &\simeq  \frac{4\pi}{3}\left(\frac{2\pi}{k_d} \right)^3 R_{\rm eq}^3 \bar{\rho}(R_{\rm eq}) x^{-3} \\
	&\simeq 3.5\cdot 10^{-10} M_\odot A_d^{-3} \left(\frac{10^{12}~\GeV}{f_a}\right)^6 \left(\frac{10^{-6}~\eV}{m_a}\right)^3 x^{-3} ~.
\end{aligned}
\eeq
If $x\sim \mathcal{O}(10)$, $M_d$ varies from roughly $10^{-22} M_\odot$ to $10^2 M_\odot$ over the allowed parameter space in Figure~\ref{fig:signal}. Larger mass substructure will form at larger scales from the $k^3$ part of the initial density power spectrum, as these modes grow and collapse during matter domination, however the resulting clumps have lower density and are more likely to be subsequently destroyed by e.g. tidal disruption  \cite{Gorghetto:2022sue}.

\section{Discussion and future directions} \label{sec:conclusions}

We have shown that it is plausible that a post-inflationary axion-like-particle with $N>1$ can comprise the full dark matter abundance with values of 
$g_{a\gamma\gamma}$ that are fairly easily discoverable by several proposed haloscopes and are also relevant for searches using astrophysical observations, see Figure~\ref{fig:signal}.\footnote{Among others, a post-inflationary $N>1$ ALP could be discovered by the experiments TOORAD \cite{Marsh:2018dlj}, BREAD \cite{BREAD:2021tpx}, BRASS \cite{brass}, ORGAN \cite{McAllister:2017lkb}, MADMAX \cite{Caldwell:2016dcw}, ALPHA \cite{Lawson:2019brd}, the next stages of ADMX, KLASH \cite{Alesini:2017ifp}, DM-Radio \cite{Silva-Feaver:2016qhh}, WISPLC \cite{Zhang:2021bpa}, the broadband mode of ABRACADABRA \cite{Ouellet:2018beu}, and searches with polaritons \cite{Mitridate:2020kly} (projections are not plotted for all of these). Meanwhile, parts of the parameter space are constrained by 
	observations by the telescopes VIMOS \cite{Grin:2006aw} and MUSE \cite{Regis:2020fhw} and extra-galactic background light (EBL), and the parameter space is close to  bounds from measurements of the polarisation of photons emitted from magnetic white dwarf stars \cite{Dessert:2022yqq} (`MWD').} Even larger $\tilde{g}_{a\gamma\gamma}$ are possible if the axion makes up only a fraction of the dark matter, but the discoverability in this case depends on the details of a particular experiment. Meanwhile, the DM substructure that is likely to form in $N>1$ theories with $H_d\ll m_a$ could provide complementary observational prospects. For instance, the miniclusters could be detected through gravitational interactions (e.g. by searches with pulsar timing arrays \cite{Ramani:2020hdo}) or via an ALP-photons coupling (e.g. by radio signals from collisions between miniclusters and neutron stars \cite{Tkachev:2014dpa,Pshirkov:2016bjr,Buckley:2020fmh,Edwards:2020afl}).\footnote{We note that the same theory can also lead to interesting signals if the string network survives until the time when the cosmic microwave background is formed and the axion makes up only a small fraction of DM \cite{Agrawal:2020euj}.} 
	
Although large parts of our analysis have rested on analytic expectations, we have made some progress in studying the string-domain wall system with numerical simulations. In particular, we have provided evidence that $\mathcal{A}_\star$ increases proportionally to $\log(m_r/H_\star)$, see Figure~\ref{fig:arasim}. As a result, when extrapolated to the physical point ($\log(m_r/H_\star)=\mathcal{O}(100)$), $\mathcal{A}_\star$ is likely to take a value that is at least an order of magnitude larger than occurs in simulations. Such results are also directly applicable to the QCD axion, for which $\log(m_r/H_\star)\simeq65$ and $\mathcal{A}_\star/N\simeq 10$, although the  non-linear evolution of the axion waves at around $H=H_\star$ in this case could affect the domain wall dynamics.

Of course many uncertainties remain. The evolution of the string-domain wall network and the area parameter at  $H\ll H_\star$ in $N>1$ theories is uncertain, and as a result the DM abundance is not known. Similarly, the density power spectrum $\mathcal{P}_\delta$ of the relic axions is not reliably known. Although not a problem in our present work, the relation between the PQ symmetry breaking Lagrangian parameter $\epsilon$ and $H_d$ is also unknown apart from an analytic guess. This is particularly important for the QCD axion in the postinflationary scenario, because it determines whether the PQ breaking potential must be fine tuned, see e.g. \cite{Sikivie:1982qv,Chang:1998tb,Kawasaki:2014sqa,Ringwald:2015dsf,Beyer:2022ywc}.

The theory we focused on is quite minimal from a high energy perspective. All that is required is an axion potential with $N>1$ degenerate minima, a small breaking of the remaining discrete $\mathbb{Z}_N$, and the PQ symmetry to be unbroken at the end of, or after, inflation. The first of these is typical for many sources of axion mass, e.g. if it comes from a strongly coupled hidden sector 
\cite{Georgi:1998au}. The second is not only easily accommodated but is thought to be inevitable \cite{Abbott:1989jw,Banks:2010zn,Harlow:2018tng}. 
Finally, for the $f_a$ we have considered, the PQ symmetry can easily be unbroken at the end of or after inflation: either due to quantum fluctuations during inflation,  finite temperature effects after reheating, or a coupling between the inflaton and the radial mode  \cite{Kofman:1986wm,Vishniac:1986sk,Yokoyama:1989pa,Hodges:1991xs,Gorghetto:2021fsn,Bao:2022hsg}. 
Additionally, it has recently been shown that, over most of the post-inflationary ALP parameter space in Figure~\ref{fig:signal}, an ALP with a coupling to photons of the form eq.~\eqref{eq:gagg} with $E/N\simeq \mathcal{O}(1)$ is only possible if the Standard Model gauge group does not arise from a grand unified theory with a QCD axion unless the Lagrangian is fine-tuned \cite{Agrawal:2022lsp}. Consequently, as well as being relatively easily discoverable, a post-inflationary ALP could give remarkable information not only about the cosmological history of the Universe (because the PQ symmetry was unbroken after inflation) but also the high energy completion of the Standard Model.

\section*{Acknowledgements}
We thank Asimina Arvanitaki for useful discussions and Giovanni Villadoro for discussions and collaboration on related work. EH thanks Jed Thompson, David Cyncynates, and Olivier Simon for useful discussions, and acknowledges the UK Science and Technology Facilities Council for support through the Quantum Sensors for the Hidden Sector collaboration under the grant ST/T006145/1 and UK Research and Innovation Future Leader Fellowship MR/V024566/1. We acknowledge use of the University of Liverpool Barkla HPC cluster and of the CINECA Marconi Skylake partition.

\appendix

\section{More details on the pre-inflationary scenario}\label{aa:isopre}

Here we summarise some standard results about ALP dark matter in the pre-inflationary scenario, focusing on the regime where the ALP is initially close to the top of a temperature independent cosine potential, i.e.  $\pi-|\theta_0|\ll 1$ (a similar analysis can be found in e.g. \cite{Arvanitaki:2019rax}).  Such an initial condition does not correspond to a fine-tuning of a Lagrangian parameter. Rather, it is a `cosmological tuning' in that our observable Universe corresponds to a rare part of the pre-inflationary, presumably random, axion field distribution.

For $\pi-|\theta_0|\ll 1$, numerical results for the function $h$ that enters the axion relic abundance  in eq.~\eqref{eq:mis_relic} can be approximated by (to an accuracy of better than $10\%$ for $\pi-|\theta_0|\lesssim 10^{-3}$)
\beq
h(\theta_0) \simeq  \log^{\, \gamma}\left(\frac{1}{\pi-|\theta_0|} \right) ~, \qquad \gamma\simeq1.4 \, ,
\eeq
where $\gamma$ is fit numerically. Therefore, in this limit, to obtain full DM abundance requires 
\beq \label{eq:pitheta}
\pi-|\theta_0| \simeq \exp\left(-15 \left(\frac{10^{12}~\GeV}{f_a} \right)^{2\gamma} \left(\frac{10^{-6}~\eV}{m_a} \right)^{\gamma/2} \right) ~.
\eeq
Therefore for a fixed $m_a$ the axion decay constant decreases only logarithmically with $\pi-|\theta_0|$ as
\beq \label{eq:faA}
f_a \propto \frac{1}{\log(1/(\pi-|\theta_0|))^{\gamma/2}}~.
\eeq

The smallest allowed $\pi-|\theta_0|$ in eq.~\eqref{eq:faA} is determined by isocurvature bounds. Any scalar field that has mass smaller than $H_I$ during inflation obtains fluctuations during inflation that are of the order of $H_I/2\pi$. For the axion these are isocurvature and translate to fluctuations $\delta\theta_0$ in the misalignment angle of
\beq
\delta \theta_0 \simeq \frac{H_I}{2\pi f_a} ~.
\eeq
This induces isocurvature perturbations in the axion dark matter energy density $\rho_a$ at all scales of
\beq
\begin{aligned}
\frac{\delta \rho_a}{\rho_a} & \simeq \frac{1}{\rho_a}  \frac{\partial\rho_a}{\partial\theta_0} \delta\theta_0  \\ &\simeq   \frac{H_I}{2\pi f_a}  \frac{1}{\rho_a}  \frac{\partial\rho_a}{\partial\theta_0}  ~,
\end{aligned}
\eeq
and an isocurvature contribution to $\mathcal{P}_\delta$ (defined in eq.~\eqref{eq:pow}) that is independent of $k$. Observational constraints require $\delta \rho_a/\rho_a < \delta_{\rm max} \simeq 10^{-5.5}$ \cite{Planck:2018vyg}, so, in the large misalignment limit,
\beq
\frac{\gamma}{(\pi-|\theta_0|)\log(1/(\pi-|\theta_0|))} \frac{H_I}{2\pi f_a} \lesssim \delta_{\rm max} ~,
\eeq
which, in combination with eq.~\eqref{eq:pitheta}, leads to a lower bound on $f_a$ for a fixed $m_a$ if $\Omega_a=\Omega_{\rm DM}$. For $H_I\simeq \eV$ values $\pi-|\theta_0| \gtrsim  10^{-18}$  are allowed.  In principle somewhat smaller values of $H_I$ are compatible with cosmological observations provided reheating is fast enough that the Universe is radiation dominated before big bang nucleosynthesis. However, if $H_I\lesssim m_a$ our assumption that the axion starts oscillating at $H\simeq m_a$ during radiation domination fails and the expression for the relic abundance will be modified in a way that depends on the particular cosmological history. 

For completeness, we note that in the opposite regime $\theta_0 \lesssim 1$, $\rho_a \propto m_a^2 f_a^2 \theta_0^2$, so if the axion makes up all the dark matter the isocurvature constraint is $f_a \theta_0 \gtrsim H_I/(2\pi\delta_{\rm max})$. Thus, using eq.~\eqref{eq:mis_relic} with $h(\theta_0)=1$, the isocurvature bound is $ m_a \lesssim 2.5\cdot 10^{-2}~\eV \left(\eV/H_I\right)^{1/4}$.

\section{The relic abundance from strings}  \label{app:st}

Prior to the time when $H=H_\star$ the string network should emit most of the axions at momentum $x_0 H$, with $x_0$ a numerical factor that is expected to be $\simeq 10$ based on simulation results. Consequently, the axion number density at a time $t>t_\star$ as a result of emission prior to $H=H_\star$ is
\beq \label{eq:nrelic_sta}
\begin{aligned}
	n_a^{\rm st}(t) & \simeq \int_{\log t_0}^{\log t_\star} \frac{2 \xi}{x_0} \log\left(\frac{m_r}{H(t')}\right) ~ \frac{\pi f_a^2}{t'} \left(\frac{R(t')}{R(t)} \right)^3 ~d\log t' \\
	& \simeq  \frac{8 \xi_\star}{x_0} \pi m_a f_a^2  \log_\star \left(\frac{R_\star}{R(t)} \right)^3 	~,
\end{aligned}
\eeq
where we used $t_0\ll t_\star$ in the second line. Eq.~\eqref{eq:nrelic_sta} leads to an axion relic abundance given by eq.~\eqref{eq:string_relic}. 
On the other hand, between the times when $H=H_\star$ and $H=H_d$ the majority of the energy is expected to be emitted into axions that are non-relativistic or at most mildly relativistic. As a result, at a time $t_\star<t<t_d\equiv 1/(2H_d)$ the axion number density from emission from strings after $H=H_\star$ is
\beq \label{eq:nrelic_st}
\begin{aligned}
	n_a^{\rm st}(t) & \simeq \int_{\log t_\star}^{\log t} 2 \xi \log\left(\frac{m_r}{H(t')}\right) ~ \frac{\pi f_a^2}{m_a t'^2} \left(\frac{R(t')}{R(t)} \right)^3 ~d\log t' \\
	& \simeq 16 \xi_\star \pi m_a f_a^2  \log_\star \left( 1- \left(\frac{H(t)}{m_a}\right)^{1/2} \right)  \left(\frac{R_\star}{R(t)} \right)^3	~.
\end{aligned}
\eeq
Because $t_\star = 1/(2m_a)$,  from eqs.~\eqref{eq:nrelic_sta} and \eqref{eq:nrelic_st} the majority of the axions are produced soon before and soon after the time when $H=H_\star$. Comparing with eq.~\eqref{eq:naw1}, for $H_d\ll m_a$, $n_a^{\rm st}$ is much smaller than the number density emitted by domain walls, $n_a^{\rm w}$, during the same period.  In particular, this is true once $H_d/m_a$ is small enough to overcome the logarithmic enhancement in eqs.~\eqref{eq:nrelic_sta} and \eqref{eq:nrelic_st}.

\section{Simulations}  \label{app:sim}

\subsection{Details of simulations}

We solve the equations of motion of a complex scalar field $\phi$ with Lagrangian
\beq
\mathcal{L}= |\partial_\mu \phi|^2 - V(\phi) ~,
\eeq
in a Friedmann-Robertson-Walker background with scale factor $R(t)\propto t^{1/2}$, where $t$ is the cosmic time and the Hubble parameter $H=1/(2t)$. The potential $V$ is of the general form of eq.~\eqref{eq:Vphi}, and we choose the particular example
\beq \label{eq:Va}
V\left(\phi\right)= \frac{m_r^2}{2 v^2} \left( |\phi|^2- \frac{v^2}{2} \right)^2 +  \frac{m_a^2 v^2}{N^2}  \left( 1- \frac{\sqrt{2}|\phi|}{v} \cos\left(N \frac{a}{v} \right) \right) + {\tilde \epsilon}^2 m_a^2 v^2  \left( 1- \frac{\sqrt{2}|\phi|}{v} \cos\left( \frac{a}{v} \right) \right) ~,
\eeq
where $\phi= \frac{v +r}{\sqrt{2}} e^{ia/v}$. We follow \cite{Hiramatsu:2012sc} in multiplying the cosine in eq.~\eqref{eq:Va} by factor of $\sqrt{2}|\phi|/v$  to avoid numerical instabilities (this has no effect away from the string cores). The magnitude of the final, $\mathbb{Z}_N$ breaking, term is parametrised by  $\tilde{\epsilon}$, related to $\epsilon$ defined in the main text by $\tilde{\epsilon}= v\sqrt{\epsilon}/(Nm_a)$.\footnote{The choice that the minima of the $\mathbb{Z}_N$ preserving and breaking contributions to the potential coincide does not affect the dynamics.}  The EoM depend on $m_a$ only through the ratio $m_r/m_a$, so we report different values of the axion mass always in terms of this. 
We solve the EoM on a discrete lattice of constant comoving size using a standard finite difference algorithm, typically on grids containing $\sim 3000^3$ points.

The maximum ratios $m_r/H$ and $m_a/H$ that can be reached in simulations are limited by the combination of three requirements. First, the physical lattice spacing $\Delta$ must be sufficiently small that the string and domain wall cores are resolved, i.e. $\Delta m_r\lesssim 1$ and $\Delta m_a\lesssim 1$ at all times. We pick  $ \Delta m_r =1 $, which we have checked is sufficient to not introduce the systematic uncertainties \cite{Gorghetto:2018myk}. Second, while the network is present there must be at least a few Hubble patches within the simulation, i.e. $HL\gtrsim 1$ where $L$ is the physical box size. We run simulations to $HL=2$ (in Appendix~\ref{aa:sys} we show this is safe from finite volume systematics). As a result, $\log(m_r/H_\star)\lesssim \log(m_r/H)\lesssim 7$ for the grid sizes that we use. Finally, $m_a/m_r$ must be sufficiently smaller than one so that the potential in eq.~\eqref{eq:Va} leads to domain walls across which the radial mode stays close to its vacuum expectation value \cite{Fleury:2015aca} , rather than interpolating across the top of its potential (such configurations are not the physical domain walls). In practice we find $m_r/m_a \geq 5$ is sufficient, which in turn fixes $\log(m_a/H)\lesssim 5$.

We focus on the `fat string and domain wall' system in which $m_r(t) \propto m_a(t) \propto t^{-1/2}$, and only consider the evolution of the physical theory (with $m_r$ constant) in simulations of the string scaling regime (i.e. prior to the time when $H=H_\star$). In the fat string system, the  number of lattice points inside string cores and domain walls remains constant in time. As mention in the main text, the fat string and domain wall system reaches the attractor faster. Moreover, modes with physical momentum $k/R=m_r$ or $k/R=m_a$ correspond to the same comoving momentum regardless of when they are produced, so energy emitted with momentum close to the string core scale ($m_r$) does not contaminate the spectrum at smaller momenta, during the scaling regime and the domain wall evolution. 

Notice that the fat domain walls mean that the energy density emitted by the domain wall network per Hubble time evolves as  $\Gamma/H \simeq \sigma H \propto t^{-3/2}$ rather than the physical $\propto t^{-1}$, see eq.~\eqref{eq:GammaW}. However, because the energy density in axion waves redshifts $\propto R^{-4}$, the relic abundance is still dominantly produced at late times. Additionally, the energy density in domain walls $\simeq H m_a f_a^2$ still dominates that in strings $\simeq \pi f_a^2 \log(m_r/H) H^2$ when $H\ll m_a$.  

We identify strings and domain walls using the algorithms in \cite{Fleury:2015aca} and \cite{1989ApJ...347..590P,Hiramatsu:2012sc} respectively and calculate the scaling parameters $\xi$ and $\mathcal{A}$ from their definitions given in Section~\ref{sec:theory}. As in the main text, when plotting results, it is convenient to measure the time in simulations in terms of $\log(m_r/H)$  or $\log(m_a/H)$.  Similarly to \cite{Gorghetto:2018myk,Gorghetto:2020qws} $\partial \rho_a / \partial k$ can be written in terms of the Fourier transforms  $\tilde{a}({\bf k})$ and  $\tilde{\dot{a}}({\bf k})$ of the axion field $a({\bf x})$ and its time derivative  $\dot{a}({\bf x})$ respectively:
\beq \label{eq:akin}
\frac{\partial \rho_a}{\partial {k/R}} = \frac{k^2/R^2}{(2\pi)^3 \mathcal{V}} \int d\Omega\frac12\left(|\tilde{\dot{a}}({\bf k}) |^2 + \left( \frac{{\bf k}^2}{R^2} + m_a^2 \right) |\dot{a}({\bf k}) |^2 \right) ~,
\eeq
where $k$ is the magnitude of the axion comoving momentum ${\bf k}$, $\mathcal{V}$ is the simulation volume, and $\Omega$ is the solid angle.

\subsection{The scaling parameters before network destruction} \label{aa:scal}

In this Appendix we show simulation results for the string length $\xi$ and the domain wall area parameter $\mathcal{A}$ during the string scaling regime and after $H_\star$, as discussed in Section~\ref{ss:relic}. 

\begin{figure*}[t]
	\begin{center}
		\includegraphics[width=0.45\textwidth]{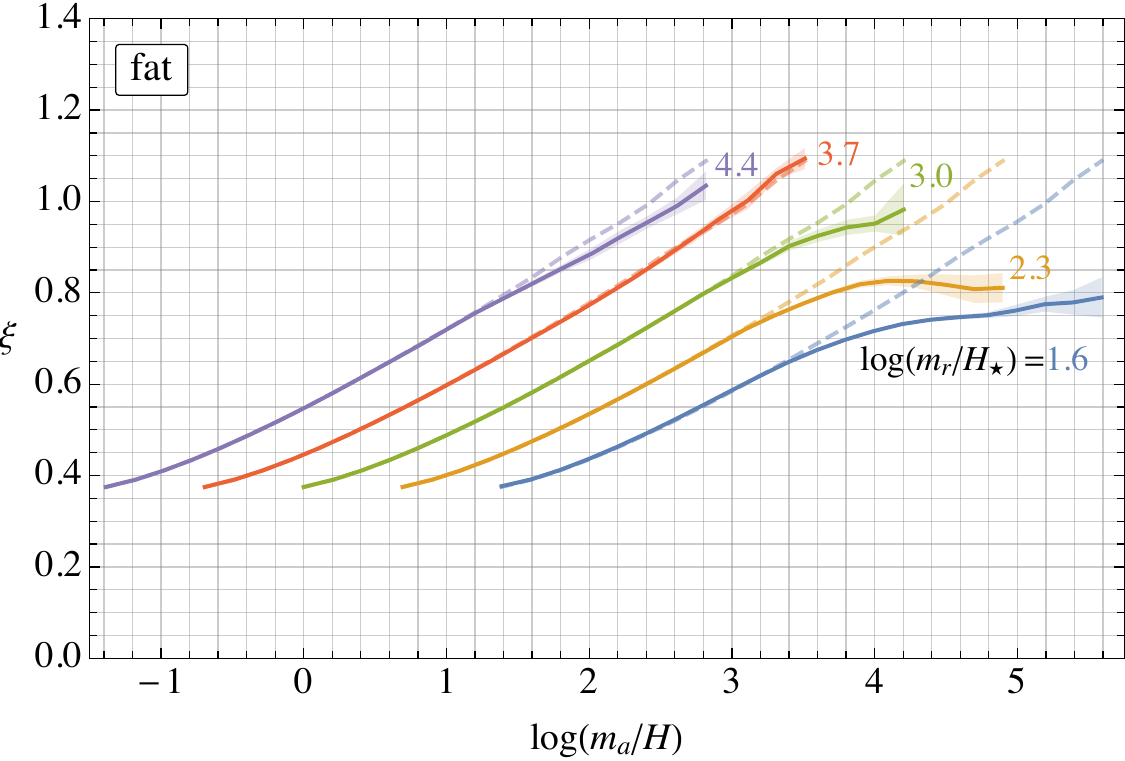} \qquad
		\includegraphics[width=0.45\textwidth]{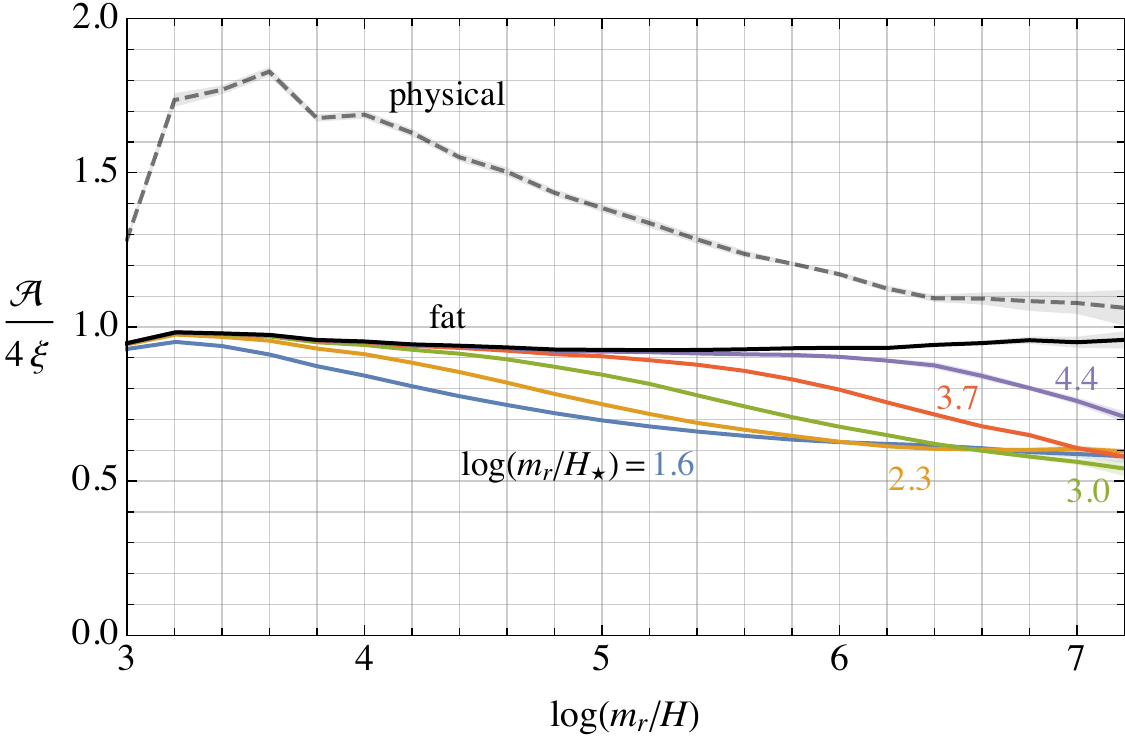} 
	\end{center}
	\caption{{\bf Left:}  The string length scaling parameters  $\xi$ as a function of time represented by $\log(m_a/H)$ for different values of the axion mass, parametrized by $\log(m_r/H_\star)= \log(m_r/m_a)$. We set the number of domain wall to $N=4$ and no explicit $\mathbb{Z}_N$ symmetry breaking and study the fat string system. Statistical uncertainties from multiple simulation runs are shaded. Dashed lines show the evolution of $\xi$ with $m_a=0$. While $\log(m_a/H)\lesssim 3.5$, $\xi$ grows as in the $m_a=0$ case unaffected by the axion mass 
		and subsequently $\xi$ deviates, affected by the axion potential and the (stable) network of domain walls that this leads to. 
		{\bf Right:}	The ratio $\mathcal{A}/\xi$ for the fat and physical string systems with $m_a=0$ and $N=4$ and, for the fat system, also with non-zero $m_a$, again parameterised by $\log(m_r/H_\star)= \log(m_r/m_a)$, and with no explicit $\mathbb{Z}_N$ symmetry breaking.
		\label{fig:scal}}
\end{figure*} 

In Figure~\ref{fig:scal} (left) we plot the evolution of $\xi$ in the fat system for different values of the axion mass, normalised as $m_a/m_r$ (i.e. different $\log(m_r/H_\star)$) with $\tilde{\epsilon}=0$ (meaning no $\mathbb{Z}_N$ breaking) and $N=4$. This is analogous to the plot of $\mathcal{A}$ in Figure~\ref{fig:arasim} (left). For comparison, we also show the evolution in the absence of the axion mass, corresponding to the string scaling regime. For sufficiently small $\log(m_a/H)$, $\xi$ has  the well-known growth proportional to $\log(m_r/H)$ \cite{Fleury:2015aca,Klaer:2017ond,Gorghetto:2018myk,Klaer:2019fxc}.  $\xi$ deviates  from the $m_a=0$ evolution at $\log(m_r/H_\star)\gtrsim 3$; indicating that the domain walls that bound the strings at such times are affecting the dynamics.  We do not attempt to extrapolate $\xi$'s evolution to large $\log(m_a/H)$ and $\log(m_a/H_\star)$.   
In Figure~\ref{fig:scal} (right) we plot the ratio $\mathcal{A}/\xi$ as a function of $\log(m_r/H)$ for the fat and physical string systems with $m_a=0$ (black lines) and also with $m_a\neq 0$ (coloured lines) for the fat strings, all with $N=4$. For fat strings this ratio seems approximately constant while $m_a\ll H$. Meanwhile, for the physical system with $m_a=0$ the evolution is less clear but is consistent with a slow approach to a constant ratio. The ratio in simulations with $m_a\neq 0$ deviates from the $m_a=0$ evolution.

Finally, we give more details on the fit of $\mathcal{A}$ prior to the time when $H=H_\star$, discussed  in the main text. For the fat string system, the form in eq.~\eqref{eq:A_linear} fits the data well, with $c_1$ consistent regardless of the $\log(m_r/H)$ in the range $3\div 5.5$ at which the fit is started.  A quadratic ansatz of the kind $\mathcal{A}(t)=c_2\log^2+c_1\log+c_0+c_{-1}\log^{-1}+c_{-2}\log^{-2}$ results in $c_2=-0.002\pm .02$ compatible with zero but with a large uncertainty. The linear form in eq.~\eqref{eq:A_linear} also fits the physical data well, but the best fit $c_1$ varies depending on the  $\log(m_r/H)$ from which the fit is started. Considering the change in $c_1$ for initial $\log(m_r/H)$ between $3$ and $5$, we estimate $c_1=0.36(10)$. We also note that in the physical case a quadratic fit of the previous form gives $c_2\simeq 0.15(10)$, slightly incompatible with zero.

\subsection{Systematic uncertainties} \label{aa:sys}

\begin{figure*}[t]
	\begin{center}
		\includegraphics[width=0.55\textwidth]{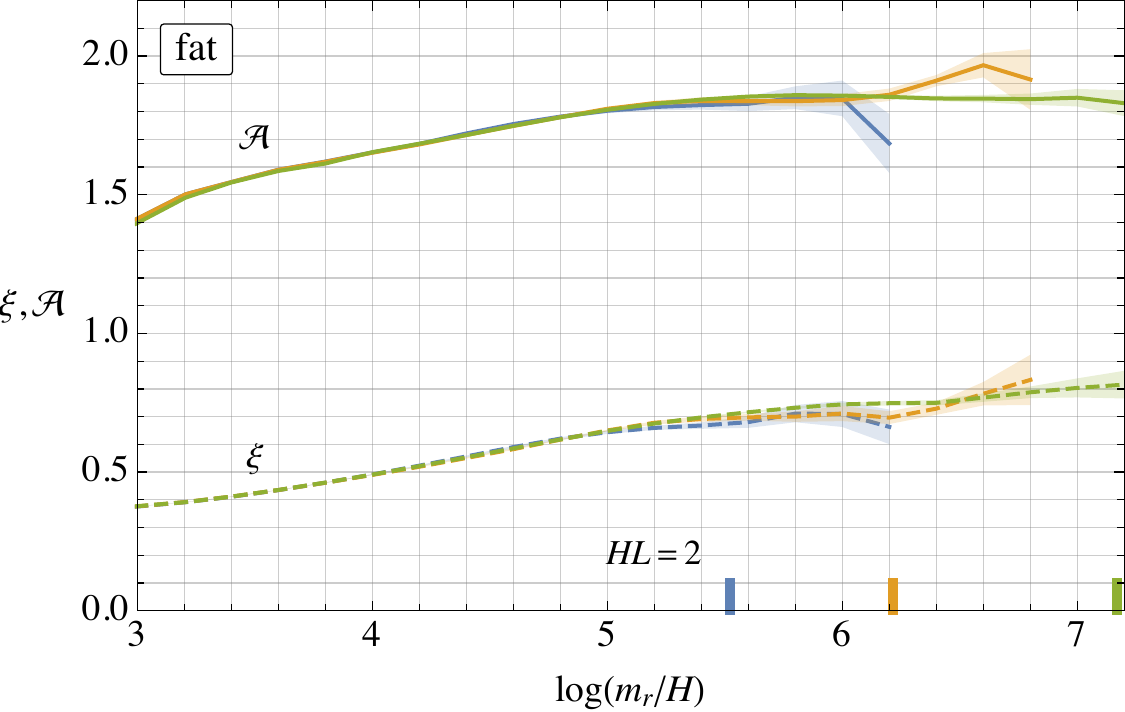} 
	\end{center}
	\caption{Comparison between the domain wall area parameter $\mathcal{A}$ and string length parameter $\xi$ as a function of time from simulations in different box sizes, but which are otherwise identical (with $m_r/m_a=5$, $\tilde{\epsilon}=0$ and $N=4$). This different simulations reach $HL=2$ at different times (indicated on the lower axis). The agreement between each simulation and the largest simulation (green), which contains many Hubble patches at those points, indicates that $\mathcal{A}$ and $\xi$ are unaffected by finite volume systematics until at least $HL=2$.
		\label{fig:HL}}
\end{figure*}

Systematic uncertainties can arise in simulations in several ways:  the finite lattice spacing, the finite volume, and the finite timestep. We have checked that our choices of simulation parameters are sufficient for all the observables that we extract from simulations to be unaffected. Since the analysis is very similar to that in \cite{Gorghetto:2018myk}, in this Appendix we simply demonstrate that our choice to end simulations when $HL=2$ is sufficient for finite volume simulations to not affect $\mathcal{A}$ (this is particularly important because the hint of a decrease in $\mathcal{A}$ in Figure~\ref{fig:arasim} right happens at the final times). To do so,  in Figure~\ref{fig:HL} we plot $\mathcal{A}$ and $\xi$ as a function of time for identical theories, but in boxes of different sizes, so that $HL=2$ happens at different $\log(m_r/H)$. After averaging over multiple runs, we see that there is no deviation in the mean $\mathcal{A}$ and $\xi$ in the smaller simulations until well after  they have $HL=2$. We have checked this remains the case regardless of $m_a/m_r$, $N$ and $\tilde{\epsilon}$.

\bibliography{references}
\bibliographystyle{utphys}

\end{document}